\documentclass[11pt,a4paper]{article}

\usepackage[utf8]{inputenc}
\usepackage[T1]{fontenc}
\usepackage{amsmath,amssymb,amsfonts,bm}
\usepackage{graphicx}
\usepackage[margin=2.5cm]{geometry}
\usepackage{setspace}
\usepackage{lineno}
\usepackage{cite}
\usepackage{xcolor}
\usepackage{soul}
\colorlet{lightyellow}{yellow!30}
\sethlcolor{lightyellow}
\usepackage{xr}
\usepackage{hyperref}
\usepackage{doi}
\hypersetup{
    colorlinks=true,
    linkcolor=blue,
    citecolor=blue,
    urlcolor=blue
}

\newcommand{\kk}{\mathbf{k}}
\newcommand{\hh}{\mathbf{h}}
\newcommand{\zhat}{\hat{\mathbf z}}

\newcommand{\uhat}{\hat{\mathbf u}}
\newcommand{\phat}{\hat{\mathbf p}}
\newcommand{\Nhat}{\hat{\mathbf N}}
\newcommand{\bsig}{\bm\sigma}
\newcommand{\MM}{\mathbf{M}}

\def \usach {Departamento de F\'isica, CEDENNA, Universidad de Santiago de Chile, 9170124, Santiago, Chile.}
\def \hcl {Hitachi Cambridge Laboratory, J. J. Thomson Avenue, CB3 0HE, Cambridge, United Kingdom.}

\title{A spin-bond theory unifying non-relativistic spin splitting and emergent spin-orbit textures}

\author{
S.\ Allende$^{1}$ and Rubén M.\ Otxoa$^{2}$ \\[1em]
\normalsize{$^{1}$\usach}\\
\normalsize{$^{2}$\hcl}\\[1em]
}

\date{\today}

\begin{document}

\maketitle

\begin{abstract}
{Magnetic order with vanishing net magnetization can produce non-relativistic spin-split bands, broadly categorized into even-parity altermagnets and odd-parity \(p\)-wave magnets. Here, we introduce a spin-bond theory that unifies these seemingly distinct phenomena into a single algebraic framework. We demonstrate that non-relativistic spin textures are fundamentally governed by two components of the electronic bond: unitary spin phases and Hermitian spin amplitudes. The unitary sector generates odd-parity p-wave and emergent spin-orbit-like textures, while the Hermitian sector generates even-parity spin fields, including the uniform \(\Gamma\)-split and bond-structured altermagnetic limits. Beyond unifying known phases, our theory uncovers a mixed non-commuting regime that emerges when the unitary and Hermitian sectors fail to commute, revealing an underlying non-commuting spin-bond structure. This regime generates a non-coplanar spin texture characterized by an even-in-momentum transverse spin polarization, providing a direct spectroscopic fingerprint for spin- and angle-resolved photoemission spectroscopy. Furthermore, we establish that this synthetic spin-orbit coupling can be dynamically tuned by geometrically controlling the non-commutation of the bond sectors. By providing a microscopic foundation for such tuning, our theory paves the way for advanced applications, including field-free spin qubits.}
\end{abstract}


\label{sec:introduction}

Magnetic order with vanishing net magnetization has historically been associated with spin-degenerate electronic bands, as is the case for conventional collinear antiferromagnets protected by combined \(\mathcal{PT}\) symmetry. However, recent discoveries have demonstrated that this degeneracy is not generic. Altermagnets provide one route beyond this limitation: they are compensated collinear magnets that exhibit momentum-dependent spin splitting with even-parity (\(d\)-, \(g\)-, or \(i\)-wave) form factors, entirely in the absence of relativistic spin--orbit coupling (SOC)~\cite{Smejkal2022a,Smejkal2022b,Mazin2022,Smejkal2020,Hayami2019,Ahn2019}. Odd-parity magnets, such as \(p\)-wave magnets, provide a second route, where non-relativistic spin splitting is odd in momentum and originates from compensated non-collinear magnetic moments~\cite{Hellenes2024,Mitscherling2026}. In both regimes, compensated magnetic order generates a robust spin splitting without relying on macroscopic magnetization or relativistic effects.

Despite the rapid expansion of this landscape~\cite{Bhowal2024,JaeschkeUbiergo2025,Karetta2026,Yu2025}, it remains unclear whether a single microscopic principle unifies these distinct symmetry classes. Establishing such a connection goes beyond fundamental magnetism, providing a direct mechanism to engineer momentum-dependent spin splitting without relying on macroscopic magnetization or relativistic spin-orbit coupling.

In this work, we establish this connection by introducing a spin-bond theory of non-relativistic spin splitting. We reveal that the central physical quantity governing these phenomena is the spin-dependent bond operator—the \(2\times2\) matrix in spin space that an electron experiences when hopping between lattice sites. Mapping the local exchange interactions onto the electronic bonds, we find that their physical information content depends on both their polar character and spatial pattern. A frame-induced unitary dressing common to every bond is removable, whereas a common Hermitian dressing is generally physical because it changes the spin-dependent hopping amplitudes.

Our theoretical approach is based on the polar decomposition of the spin-bond matrix. Just as a complex number factorizes into a modulus and a phase, any invertible bond matrix factorizes into a unitary part (carrying spin-dependent phases) and a positive Hermitian part (carrying spin-dependent amplitudes):
\begin{equation}
    T_\delta=U_\delta P_\delta ,
    \label{eq:polar_intro}
\end{equation}
with \(U_\delta\) unitary and \(P_\delta=P_\delta^\dagger>0\) Hermitian positive. This algebraic split dictates the momentum parity of the spin field. The unitary phases generate odd-in-momentum spin textures, realizing \(p\)-wave magnets along with exchange-driven synthetic SOC (such as Rashba, Dresselhaus, and out-of-plane-like textures). Conversely, the Hermitian amplitudes generate even-in-momentum textures. A uniform Hermitian pattern can produce a \(\Gamma\)-split compensated spin splitting, while bond-structured Hermitian patterns capture altermagnetism as a natural limit. Thus, the familiar even and odd non-relativistic spin-splitting classes emerge as the two fundamental algebraic sectors of a single microscopic bond operator.

For reciprocal bipartite lattices, the theory is exactly solvable. The Hamiltonian factorizes into sublattice-parity sectors \(s=\pm1\):
\begin{equation}
    H_s(\kk)= -s\,h_0(\kk)\,\sigma_0+\left[\Delta\zhat-s\,\hh(\kk)\right]\cdot\bsig,
    \label{eq:Hs_intro}
\end{equation}
revealing that the resulting spin texture is strictly governed by the competition between the uniform exchange field \(\Delta\zhat\) and the momentum-dependent spin-bond field \(\hh(\kk)\). This framework makes parity-locking transparent: unitary links yield sine form factors and odd spin fields, while Hermitian links yield cosine form factors and even spin fields. The same construction serves as a design principle for non-relativistic exchange-SOC, where varying the unitary axes continuously interpolates between magnetic classes, placing recent symmetry- and strain-controlled altermagnetic transitions~\cite{Karetta2026,Amin2024,Zhou2025} in a rigorous bond-level language.

Our theory uncovers a third, mixed non-commuting regime that appears
when the unitary and Hermitian factors are simultaneously present and
fail to commute, $[U_\delta,P_\delta]\neq0$. This regime reveals an
underlying non-commuting spin-bond structure. In this non-commuting regime, the bond can no longer be diagonalized along a single spin axis, generating a transverse spin component along \(\hat{\mathbf u}_\delta\times\hat{\mathbf p}_\delta\). In the minimal model, this non-commuting mixing produces an even-in-momentum transverse spin polarization:
\begin{equation}
    \langle s_\perp\rangle_{\rm even}(\kk)\neq0 ,
\end{equation}
 a unique combination that is strictly forbidden in pure altermagnets, pure \(p\)-wave magnets, and standard Rashba/Dresselhaus textures. This unique feature serves as a direct experimental fingerprint for non-commuting spin-bond magnetism, which can be readily mapped using Spin- and Angle-Resolved Photoemission Spectroscopy (spin-ARPES)~\cite{Dil2009,Dil2019,Krempasky2024,Reimers2024}.

This non-relativistic spin-splitting has direct implications for scalable spin-based quantum computing ~\cite{AbadilloUriel2026,Kirczenow2026,VosoughiNia2025,Steinacker2025}. Recent proposals for gate-defined spin qubits~\cite{AbadilloUriel2026,Kirczenow2026,Vakili2026}, controlled entanglement rotation~\cite{Kulig2024}, and parity-protected superconducting circuits (e.g., 'altermons'~\cite{VosoughiNia2025}) highlight the power of intrinsic momentum-dependent spin splitting. Specifically, this splitting enables magnetic-field-free, all-electrical qubit control via Electric Dipole Spin Resonance (EDSR)~\cite{Golovach2006,Nowack2007,NadjPerge2010}, as well as advanced quantum sensing protocols~\cite{Sun2026}. However, current theoretical models rely largely on macroscopic crystal symmetries. Crucially, we demonstrate that an emergent synthetic SOC arises from the non-commutation of the spin-dependent bond operators rather than relativistic effects. Due to the fact that this interaction is rooted in the lattice geometry and local exchange interactions rather than fixed atomic properties, it is fundamentally tunable via strain or gating. Unlike the persistent SOC in conventional materials that permanently exposes spins to environmental electrical fluctuations (charge noise)~\cite{Borhani2006,Paladino2014,Yoneda2018,Burkard2023}, our synthetic SOC can be activated for rapid spin manipulation and subsequently turned off on-demand to protect quantum coherence. Therefore, our unified microscopic theory of compensated magnetism provides the exact lattice-level blueprint needed to locally engineer and optimize these highly coherent quantum states~\cite{Ouassou2023,Beenakker2023,Giil2024,Lu2024,Banerjee2024}.
\section*{Results}

\subsection*{Spin-bond Hamiltonian}

\begin{figure}[th]
\centering
\includegraphics[width=1\columnwidth]{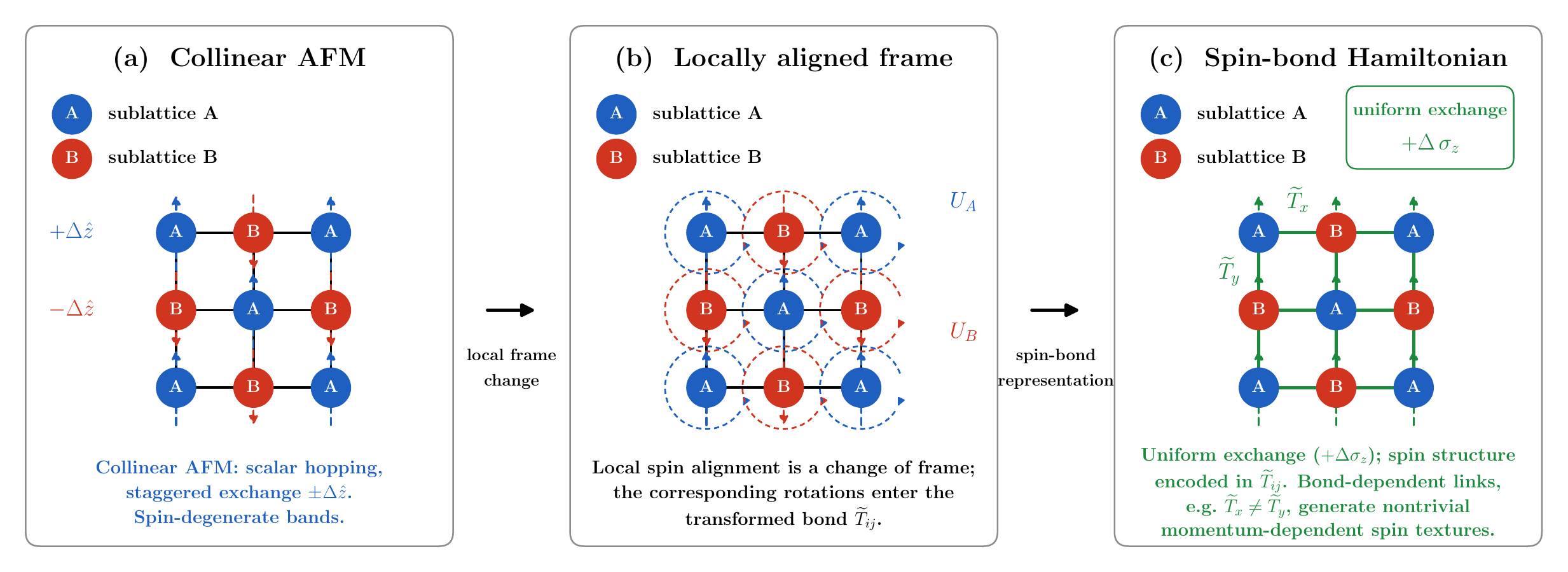}
\caption{\textbf{From a collinear antiferromagnet to a spin-bond Hamiltonian.}
(a) The reference Néel state: opposite sublattices carry opposite exchange fields $\pm\Delta\hat{\mathbf z}$ and the hopping is a spin-neutral scalar. (b) A local spin rotation $U_i$ aligns the exchange on every site, at the price of dressing the hopping, $\widetilde T_{ij}=U_i^\dagger T_{ij}U_j$. (c) The resulting spin-bond Hamiltonian: a uniform exchange $\Delta\sigma_z$ and bond-dependent links $\widetilde T_x\neq\widetilde T_y$ that carry the magnetic bond structure. A uniform unitary dressing generated solely by the local spin-frame transformation is removable, whereas a uniform Hermitian dressing can remain physical. Bond-dependent links generate nontrivial momentum-dependent spin textures.}\label{fig:afm_to_spinbond}
\end{figure}

We begin from a collinear antiferromagnet, where the hopping is a spin-neutral scalar, and ask what new physics appears once the hopping itself carries spin structure. To isolate the physics associated with a spin-dependent electronic hopping, we start from the simplest reference system: a collinear antiferromagnet on a bipartite lattice with sublattices \(A\) and \(B\) (see Fig.~\ref{fig:afm_to_spinbond}).

For definiteness we take a square lattice and nearest-neighbour hopping between opposite sublattices, although the construction is not restricted to this geometry. The real-space Hamiltonian is:
\begin{equation}
H_0=-\sum_{\langle i\in A,j\in B\rangle,\sigma}\left(t_{ij}\, c_{i\sigma}^\dagger c_{j\sigma}+ \text{h.c.}\right)+\sum_{i\in A}\Delta\, c_i^\dagger\sigma_z c_i -\sum_{j\in B}\Delta\, c_j^\dagger\sigma_z c_j .
\label{eq:H0_real}
\end{equation}

Here \(c_{i\sigma}^\dagger\) and \(c_{i\sigma}\) create and annihilate an electron with spin \(\sigma\) on site \(i\), \(t_{ij}\) is the hopping amplitude, and \(\Delta\hat{\mathbf z}\) is the local exchange field on sublattice \(A\), while \(-\Delta\hat{\mathbf z}\) is the local exchange field on sublattice \(B\). Thus the itinerant electrons move in a staggered spin potential: their spin tends to align with opposite local magnetic moments on the two sublattices.

Fourier transforming to momentum space, with the spinor
\begin{equation}
\Psi_{\mathbf k}=\left(c_{A\uparrow}(\mathbf k), c_{A\downarrow}(\mathbf k), c_{B\uparrow}(\mathbf k), c_{B\downarrow}(\mathbf k) \right)^T ,
\end{equation}
one obtains (Supplementary Note A)
\begin{equation}
H_0(\mathbf k)=-(\tau_+\otimes\sigma_0)g(\mathbf k)-(\tau_-\otimes\sigma_0)g^*(\mathbf k)
+\Delta(\tau_z\otimes\sigma_z),
\label{eq:H0_k}
\end{equation}
where
\begin{equation}
g(\mathbf k)=\sum_\delta t_\delta e^{i\mathbf k\cdot\delta}.
\label{eq:gk}
\end{equation}
The Pauli matrices \(\tau_i\) act in sublattice space, while \(\sigma_i\) act in
real spin space. At this stage nothing unconventional has been introduced. This
is the standard collinear antiferromagnet. Its spectrum is
\begin{equation}
E_\pm(\mathbf k)=\pm\sqrt{|g(\mathbf k)|^2+\Delta^2},
\end{equation}
with a twofold spin degeneracy protected by the combined symmetry
\(\mathcal{PT}\). There is no spin-orbit coupling, no crystallographic spin
texture, and no non-relativistic spin splitting.
To lift this spin degeneracy and describe altermagnetism, we must introduce the spin structure into the bonds themselves. The key point is that an electron moving through a magnetic crystal feels not only the local exchange field at each site, but also the spin-dependent matrix associated with a bond. We therefore replace the scalar hopping  \(t_{ij}\sigma_0\) by a spin-dependent bond operator \(T_{ij}\), and write:
\begin{equation}
H=-\sum_{\langle ij\rangle}\left(c_i^\dagger T_{ij}c_j+\text{h.c.}\right)+\sum_i\Delta\,c_i^\dagger
\hat{\mathbf n}_i\cdot\boldsymbol\sigma c_i .
\label{eq:H_general_real}
\end{equation}
The operator \(T_{ij}\) is a \(2\times2\) matrix in spin space. It may arise from
orbital downfolding, inequivalent hopping paths, non-collinear local moments,
interfacial environments, or any microscopic mechanism that makes the electronic
hopping spin dependent. This is the physical input of the theory. For simplicity
we take the exchange magnitude \(\Delta\) uniform, so that the local moments
differ only in their orientation \(\hat{\mathbf n}_i\).
It is useful to describe the system in a local spin frame in which the exchange field is aligned at every site. Let \(U_i\in SU(2)\) be the spin rotation that takes the local moment direction \(\hat{\mathbf n}_i\) to a common axis
\begin{equation}
U_i^\dagger\left(\hat{\mathbf n}_i\cdot\boldsymbol\sigma\right)U_i=\sigma_z .
\label{eq:local_alignment}
\end{equation}
In this local frame the exchange becomes uniform, \(\Delta\,\sigma_z\), while the
hopping becomes
\begin{equation}
T_{ij}\longrightarrow\widetilde T_{ij}=U_i^\dagger T_{ij}U_j .
\label{eq:Tcov}
\end{equation}
This bond operator is the central object of the theory: the exchange field has been made locally simple, but the price is that the hopping now carries the magnetic and crystallographic information.
This framework clearly isolates the origin of the spin splitting. The local spin rotation simplifies the exchange field to a uniform, momentum-independent term \(\Delta\,\sigma_z\), which in the reference antiferromagnet leaves the bands spin-degenerate. Any non-relativistic, momentum-dependent spin splitting, altermagnetic (even in \(\mathbf k\)) or \(p\)-wave (odd in \(\mathbf k\)), is therefore forced to originate in the bond operator \(\widetilde T_{ij}\) rather than in the exchange: as an electron hops from one site to the next, its spin is rotated and reweighted according to the local magnetic environment of that bond.

A unitary dressing originating solely from a change of local spin frame is a gauge artefact and leaves the spectrum unchanged. A uniform Hermitian factor, however, is generally physical because it changes the spin-dependent hopping amplitudes and can already produce an even-parity spin splitting. In the square-lattice realization considered below, this uniform Hermitian limit corresponds to the \(\Gamma\)-split case. Bond-dependent spin structure then generates the nontrivial \(d\)-wave, \(p\)-wave, and mixed momentum-space textures considered below. The explicit frame-induced unitary construction is given in Supplementary Note B.

\subsection*{Spin-bond Theory}
\label{sec:spinbond_theory}

Having identified the bond operator as the only place where genuine spin structure can reside, we now build the theory around it.
 
We can now formally construct the physical spin-bond operator: for a translationally invariant spin-bond pattern, the Bloch Hamiltonian becomes
\begin{equation}
H(\mathbf k)=\Delta(\tau_0\otimes\sigma_z)-(\tau_+\otimes S(\mathbf k))-(\tau_-\otimes S^\dagger(\mathbf k)),
\label{eq:H_spinbond}
\end{equation}
with
\begin{equation}
S(\mathbf k)=\sum_\delta T_\delta e^{i\mathbf k\cdot\delta}.
\label{eq:S_spinbond}
\end{equation}
From here on \(T_\delta\) denotes the bond operator
\(\widetilde T_{i,i+\delta}=U_i^\dagger T_{i,i+\delta}U_{i+\delta}\) of
Eq.~\eqref{eq:Tcov} for a translationally invariant pattern; we drop the tilde to
lighten the notation. Here $\delta$ represents the directed $A\to B$ bonds entering
$S(\mathbf{k})$. The exchange field is now featureless; all the nontrivial
magnetic information is carried by the spin-bond hopping matrix \(S(\mathbf k)\).

A fundamental insight emerges from the polar decomposition of the connection, since just as a complex number factorizes into modulus and phase, any invertible bond matrix factorizes into a positive Hermitian part and a unitary part,
\begin{equation}
T_\delta=U_\delta P_\delta ,
\label{eq:polar_basic}
\end{equation}
where \(P_\delta=P_\delta^\dagger>0\) plays the role of the modulus (a
spin-dependent \emph{amplitude}) and \(U_\delta\) plays the role of the phase (a
spin-dependent \emph{phase}). Each factor alone already yields nontrivial
physics: a purely unitary bond produces an odd-parity \(p\)-wave / emergent-SOC
texture, while a purely Hermitian bond produces an even-parity spin field. Altermagnetism is a bond-structured realization of the Hermitian amplitude sector. The new feature relative to a scalar is that modulus and phase are now matrices; when both are present and fail to commute, \([U_\delta,P_{\delta}]\neq 0\), the bond enters a mixed, non-commuting regime that is neither a pure altermagnet nor a pure \(p\)-wave magnet.

In full generality, an invertible \(2\times2\) spin-bond matrix may be written as
\begin{equation}
T_\delta=t_\delta e^{i\phi_\delta}e^{i\alpha_\delta \hat{\mathbf u}_\delta\cdot\boldsymbol\sigma}
e^{\beta_\delta \hat{\mathbf p}_\delta\cdot\boldsymbol\sigma}, \qquad t_\delta>0.
\label{eq:polar_general}
\end{equation}
Here \(t_\delta>0\) is the spin-independent hopping amplitude and \(\phi_\delta\)
a spin-independent Peierls phase. In the unitary factor, \(\hat{\mathbf u}_\delta\)
is a unit vector and \(\alpha_\delta\) a real parameter controlling the strength
of the spin-dependent phase; in the Hermitian factor, \(\hat{\mathbf p}_\delta\)
is a unit vector and \(\beta_\delta\) a real parameter controlling the strength of
the spin-dependent amplitude. The unit vectors \(\hat{\mathbf u}_\delta\) and
\(\hat{\mathbf p}_\delta\) set the spin axes of the two sectors, and their relative
orientation governs the mixed non-commuting regime. In the
following we set \(t_\delta\) real and \(\phi_\delta=0\) to isolate the
spin-dependent physics. A minimal microscopic realization in which
the ordered product $T_\delta=t_\delta U_\delta P_\delta$ emerges from
virtual hopping through an exchange-split intermediate orbital is
derived in Supplementary Note C.

This formalism naturally classifies the connections into three distinct categories---phase, amplitude, and mixed links---where the separation into unitary and Hermitian components carries a direct physical meaning: \(U_\delta\) governs spin-dependent \emph{phases}, whereas \(P_\delta\) governs spin-dependent \emph{amplitudes}. A purely unitary bond gives an odd-parity p-wave / emergent-SOC field, while a purely Hermitian bond gives an even-parity spin field. Within the Hermitian sector, a uniform pattern can produce a \(\Gamma\)-split limit~\cite{Yuan2024}, whereas bond-structured patterns can realize altermagnetic textures. A third possibility arises when
both are present and their spin axes \(\hat{\mathbf u}\) and \(\hat{\mathbf p}\)
are non-collinear, so that they fail to commute,
\([U_\delta,P_{\delta}]\neq0\); this mixed, non-commuting regime is explored below.

All momentum dependence of the spin splitting resides in the spin-bond matrix
\(S(\mathbf k)=\sum_\delta T_\delta e^{i\mathbf k\cdot\delta}\), which we decompose
into a charge part and a spin part,
\begin{equation}
S(\mathbf k) = h_0(\mathbf k)\,\sigma_0+\mathbf h(\mathbf k)\cdot\boldsymbol\sigma .
\label{eq:S_decomposition}
\end{equation}
The exchange enters the Bloch Hamiltonian in Eq.~\eqref{eq:H_spinbond} only through the uniform, \(\mathbf k\)-independent term \(\Delta(\tau_0\otimes\sigma_z)\), which is even by construction and carries no spin texture; the entire momentum-dependent spin texture is therefore the vector field \(\mathbf h(\mathbf k)\). 

For the two pure polar sectors, the momentum parity is fixed by the
character of the bond,
\begin{equation}
U_\delta\ (\text{phase})
\;\Rightarrow\;
\mathbf h(-\mathbf k)=-\mathbf h(\mathbf k),
\qquad
P_\delta\ (\text{amplitude})
\;\Rightarrow\;
\mathbf h(-\mathbf k)=+\mathbf h(\mathbf k).
\label{eq:parity_sectors}
\end{equation}

For bond patterns satisfying the reciprocal condition $T_{-\delta}=T_\delta^\dagger$, the pure unitary sector obeys $U_{-\delta}=U_\delta^\dagger$, producing odd ($\sin$) form factors,
whereas in the pure Hermitian sector
$P_{-\delta}=P_\delta$, producing even ($\cos$) form factors.
When $[U_\delta,P_\delta]\neq0$, reciprocity must instead be applied
to the full bond operator $T_\delta$. The complete derivation is given
in Supplementary Note D.

Figure~\ref{fig:spinbond_classification} summarizes the resulting algebraic classification: the unitary sector generates odd-parity p-wave/exchange-SOC textures, the Hermitian sector generates even-parity \(\Gamma\)-split and altermagnetic limits, and the product \(U_\delta P_\delta\) defines the mixed non-commuting sector when \([U_\delta,P_\delta]\neq0\).

\begin{figure}[th]
    \centering
    \includegraphics[width=0.8\textwidth]{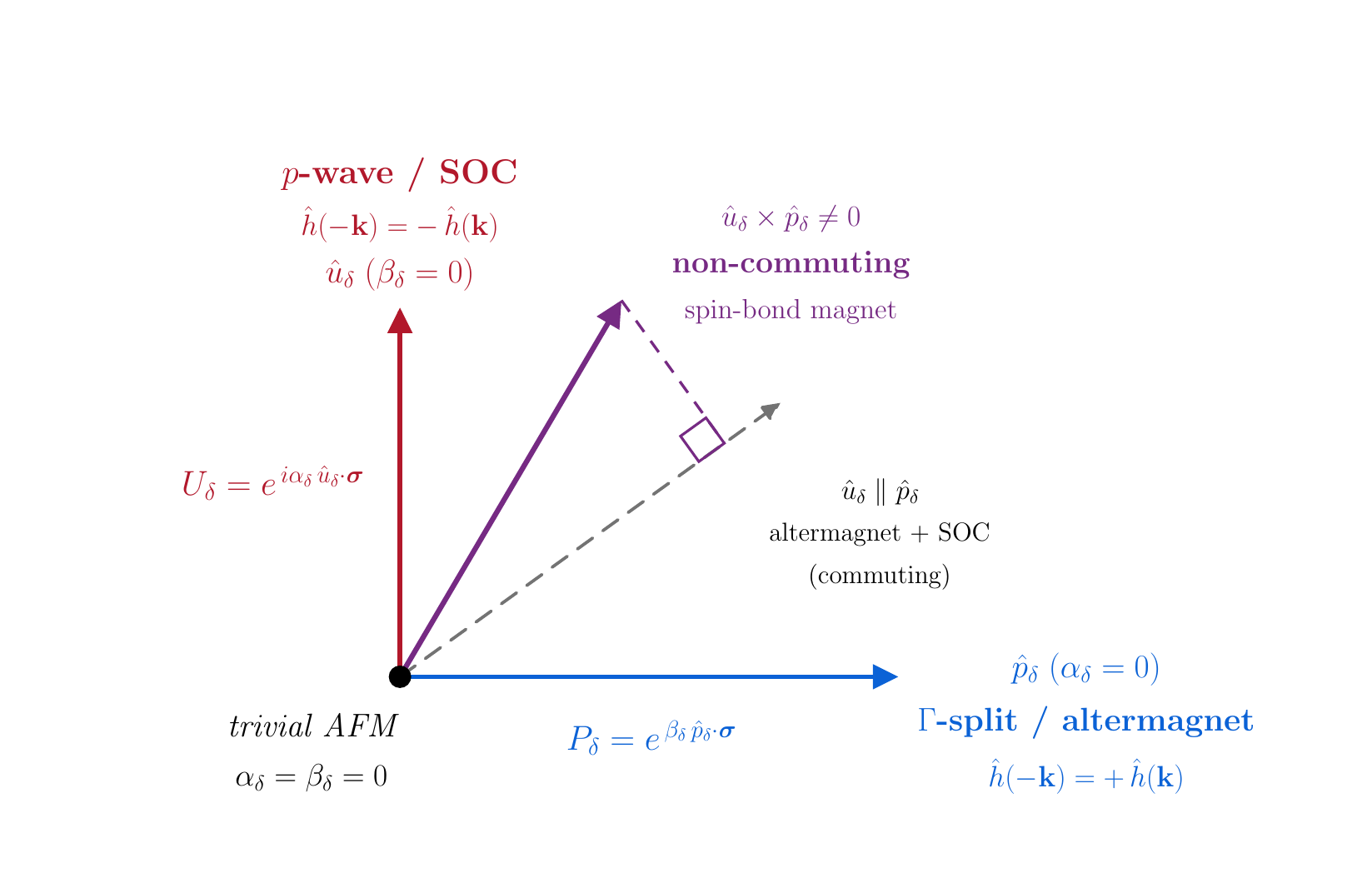}
    \caption{\textbf{Algebraic classification of spin-bond links.} The polar decomposition
    \(T_\delta=U_\delta P_\delta\) separates spin-dependent phases from
    spin-dependent amplitudes. The unitary sector \(U_\delta=e^{i\alpha_\delta\hat{\mathbf u}_\delta\cdot\boldsymbol\sigma}\)
    generates odd-parity spin fields and realizes \(p\)-wave/exchange-SOC
    textures,the Hermitian sector \(P_\delta=e^{\beta_\delta\hat{\mathbf p}_\delta\cdot\boldsymbol{\sigma}}\) generates even-parity spin fields, including the uniform \(\Gamma\)-split and bond-structured altermagnetic limits. If
    both sectors are present and
    \(\hat{\mathbf u}\parallel\hat{\mathbf p}\), the links commute and the
    texture remains coplanar. If
    \(\hat{\mathbf u}\times\hat{\mathbf p}\neq0\), the product $U_\delta P_\delta$ defines the mixed non-commuting spin-bond sector with an emergent transverse axis
    \(\hat{\mathbf u}\times\hat{\mathbf p}\). The origin
    \(\alpha=\beta=0\) corresponds to the spin-degenerate collinear AFM.
    }
    \label{fig:spinbond_classification}
\end{figure}


\subsection*{Exact solution on bipartite lattices}

Having identified the spin field \(\mathbf h(\mathbf k)\) and its parity, we now
solve the model. Under the structural conditions below the Hamiltonian factorizes
exactly, giving closed forms for the bands and the spin texture in terms of
\(\mathbf h(\mathbf k)\). To achieve this exact block factorization, we require four conditions: (i) the lattice is bipartite, with two sites per unit cell, \(A\) and \(B\). (ii) the hopping connects only opposite sublattices, at any range, with no \(A\)--\(A\) or \(B\)--\(B\) terms and (iii) the exchange is uniform and collinear, \(+\Delta\hat{\mathbf z}\) on \(A\) and \(-\Delta\hat{\mathbf z}\)
on \(B\). 
(iv) The bond pattern satisfies
$T_{-\delta}=T_\delta^\dagger$. The square lattice is the canonical realization: \(S(\mathbf k)\) is built from nearest-neighbour \(A\)--\(B\) bonds, and from farther inter-sublattice shells when richer textures are needed.

The reciprocal condition $T_{-\delta}=T_\delta^\dagger$ makes the
Bloch hopping matrix Hermitian, $S(\mathbf{k})=S^\dagger(\mathbf{k})$.
Consequently, the scalar and vector fields $h_0(\mathbf{k})$ and
$\mathbf{h}(\mathbf{k})$ in Eq.~\eqref{eq:S_decomposition} are real. With these ingredients in place, the Bloch Hamiltonian in Eq.~\eqref{eq:H_spinbond} decouples in the eigenbasis of \(\tau_x\) into two independent blocks labelled by \(s=\pm1\) (Supplementary Note E),
\begin{equation}
H_s(\mathbf k)=-s\,h_0(\mathbf k)\,\sigma_0+\left[\Delta\hat{\mathbf z}-s\,\mathbf h(\mathbf k) \right]\cdot\boldsymbol\sigma,\qquad s=\pm1.
\label{eq:Hs}
\end{equation}
Each block is a two-level spin problem in an effective field
\(\Delta\hat{\mathbf z}-s\,\mathbf h(\mathbf k)\): the uniform exchange along
\(\hat{\mathbf z}\) competes with the momentum-dependent bond field
\(\mathbf h(\mathbf k)\), with a sign that alternates between the two blocks.

This block structure allows us to readily extract the band energies and spin texture: diagonalizing Eq.~\eqref{eq:Hs} gives the bands
\begin{equation}
E_{s,\pm}(\mathbf k)=-s\,h_0(\mathbf k)\pm\left|\Delta\hat{\mathbf z}-s\,\mathbf h(\mathbf k)\right|,
\label{eq:bands}
\end{equation}
and the spin texture
\begin{equation}
\langle\mathbf s\rangle_{s,\pm}(\mathbf k)=\pm\,\hat{\mathbf m}_s(\mathbf k),\qquad\hat{\mathbf m}_s(\mathbf k)=\frac{\Delta\hat{\mathbf z}-s\,\mathbf h(\mathbf k)}{\left|\Delta\hat{\mathbf z}-s\,\mathbf h(\mathbf k)\right|}.
\label{eq:texture}
\end{equation}

Throughout, \(\mathbf h(\mathbf k)\) denotes the bond field with its magnitude, and a hat denotes a unit vector, \(\hat{\mathbf h}=\mathbf h/|\mathbf h|\). Here \(\langle\mathbf s\rangle\) is the unit spin-polarization direction --- the expectation of \(\boldsymbol\sigma\), normalized to unit length --- so that the spin points along the unit vector \(\hat{\mathbf m}_s\) set by the competition between the uniform exchange and the bond field. This is the convention used in the figures.

The bond field \(\mathbf h(\mathbf k)\) is the quantity that governs the spin texture. When \(\mathbf h(\mathbf k)=0\) the effective field reduces to \(\Delta\hat{\mathbf z}\), the texture is collinear with the exchange, and the bands are spin degenerate: this is the reference collinear antiferromagnet. A non-relativistic, momentum-dependent spin splitting appears whenever \(\mathbf h(\mathbf k)\neq0\), and its parity --- odd for a phase bond, even for an amplitude bond, Eq.~\eqref{eq:parity_sectors} is what distinguishes the \(p\)-wave and altermagnetic sectors in the bands and the spin texture.

Before turning to the mixed non-commuting regime, we show that this single construction already reproduces the relevant non-relativistic spin textures. The \(\Gamma\)-split (uniform Hermitian), altermagnetic, p-wave, Rashba, Dresselhaus, radial, and out-of-plane limits correspond to specific choices of bond amplitudes and axes, summarized in Table~\ref{tab:recipes} and derived in Supplementary Note F. These exchange-generated textures share the momentum dependence of
conventional Rashba, Dresselhaus, and Weyl spin-orbit coupling, but
their microscopic origin is different. Relativistic SOC produces a
spin-dependent hopping of the generic form $T_{\delta}^{\mathrm{rel}}
=
t_{\delta}\sigma_{0}
+
i\,\boldsymbol{\lambda}_{\delta}^{\mathrm{rel}}
\cdot\boldsymbol{\sigma}$.
Its polar decomposition contains a unitary factor with the same SU(2)
structure as $U_{\delta}$. Hence, The two
mechanisms cannot be distinguished by the momentum-space
texture alone. Relativistic SOC originates from relativistic matrix
elements of the crystal potential and vanishes formally as
$c\rightarrow\infty$, whereas the spin-bond contribution is generated
by exchange and by a bond-dependent spin structure that cannot be
gauged away. We therefore refer to the resulting fields as emergent
spin-orbit textures. The unitary
factors of both contributions combine into a single link, which we
denote by $U_{\delta}^{\mathrm{tot}}$; a relativistic contribution can
therefore rotate the effective spin axis of the unitary sector without
generating a new polar one.

\begin{table}[th]
\centering
\renewcommand{\arraystretch}{1.5}
\begin{tabular}{l|l|l|l}
class & unitary $U_\delta$ & Hermitian $P_\delta$ & spin part $\mathbf h(\kk)\cdot\boldsymbol\sigma$\\\hline
$\Gamma$-split
& $\sigma_0$
& $\hat{\bm p}=\hat{\bm z},\;\beta_x=\beta_y=\beta$
& $2t\sinh\beta(\cos k_x+\cos k_y)\sigma_z$\\
$p$-wave
& $\uhat=\zhat$, $\alpha_x{=}{+}\alpha,\ \alpha_y{=}{-}\alpha$
& $\sigma_0$
& $\lambda(\sin k_x{-}\sin k_y)\,\sigma_z$\\
$d$-altermagnet
& $\sigma_0$
& $\phat=\zhat$, $\beta_x{=}{+}\beta,\ \beta_y{=}{-}\beta$
& $2t\sinh\beta(\cos k_x{-}\cos k_y)\,\sigma_z$\\
Rashba
& $\uhat_x{=}\hat{\mathbf y},\ \uhat_y{=}{-}\hat{\mathbf x}$
& $\sigma_0$
& $\lambda(k_x\sigma_y{-}k_y\sigma_x)$\\
Dresselhaus
& $\uhat_x{=}\hat{\mathbf x},\ \uhat_y{=}{-}\hat{\mathbf y}$
& $\sigma_0$
& $\lambda(k_x\sigma_x{-}k_y\sigma_y)$\\
radial (Weyl)
& $\uhat_x{=}\hat{\mathbf x},\ \uhat_y{=}\hat{\mathbf y}$
& $\sigma_0$
& $\lambda(k_x\sigma_x{+}k_y\sigma_y)$\\
out-of-plane
& $\uhat_x{=}\zhat$
& $\sigma_0$
& $\lambda\,k_x\,\sigma_z$\\
\end{tabular}
\caption{Spin textures generated by the spin-bond construction on the square lattice. Each class follows from a choice of bond amplitudes and axes through the same field \(\mathbf h(\mathbf k)\). The Rashba, Dresselhaus, radial, and out-of-plane rows are shown to leading order in \(k\) (\(\lambda\equiv2t\sin\alpha\)), while the \(p\)-wave, \(\Gamma\)-split, and \(d\)-altermagnet forms are exact. The \(\Gamma\)-split row denotes the uniform Hermitian limit, which remains spin split at the Brillouin-zone centre. Derivations are given in Supplementary Note F}.
\label{tab:recipes}
\end{table}

\begin{figure}[htb]
  \centering
  \includegraphics[width=0.7\textwidth]{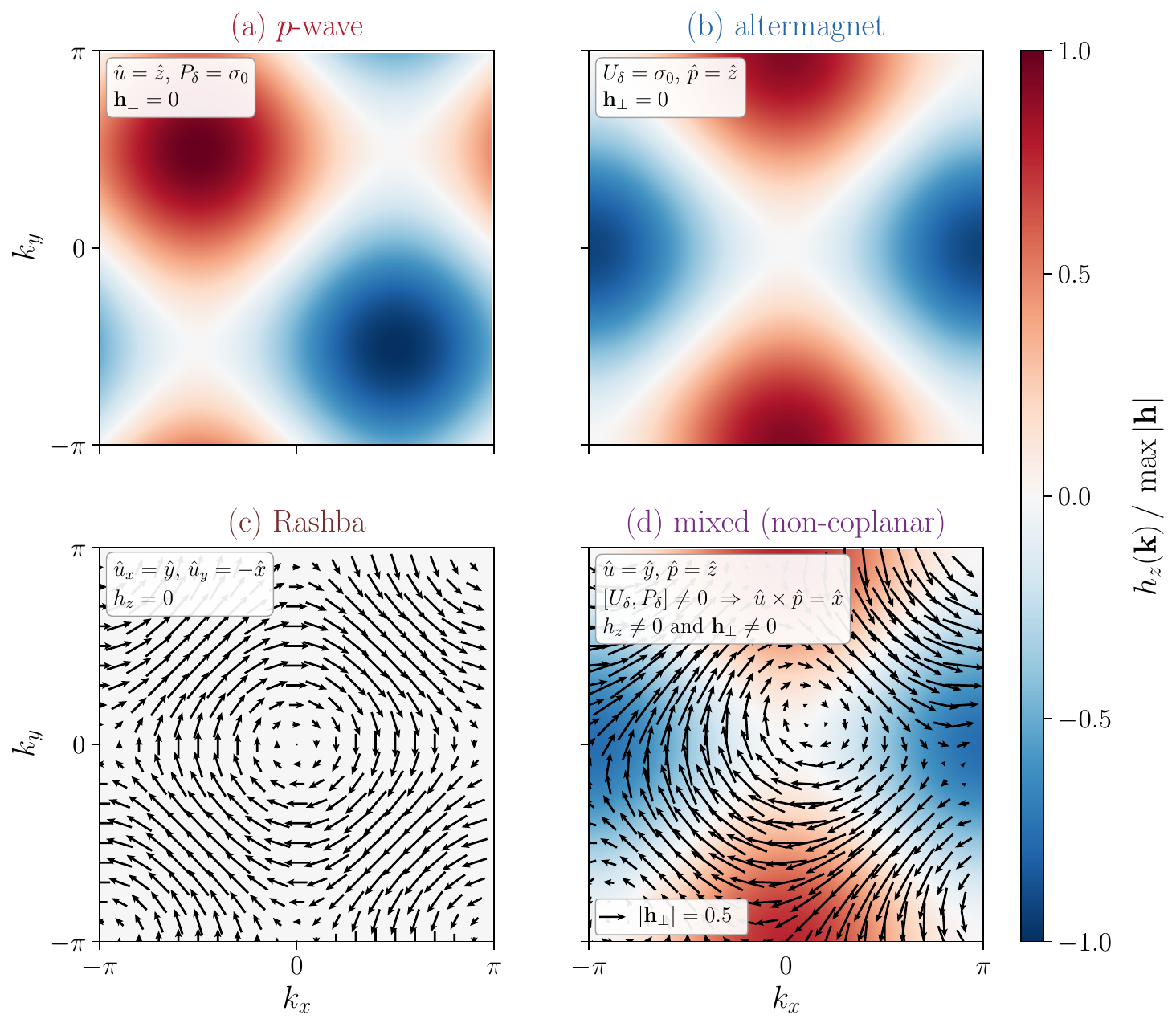}
  \caption{\label{fig:spin_textures}
  Momentum-space spin-bond textures ($t=1$, $\alpha=0.6$, $\beta=0.5$, illustrative bond parameters). Momenta are in units of the inverse lattice constant $a=1$. In every panel
  the color is the longitudinal component $h_z$, along the N\'eel axis
  $\hat{\mathbf{z}}$ of the uniform exchange $\Delta\hat{\mathbf{z}}$, and the arrows
  are the transverse component $\mathbf{h}_{\perp}=(h_x,h_y)$; both are normalized to
  the largest $|\mathbf{h}|$ over the four panels, with the arrow scale given in (d).
  The bond axes $\hat{\mathbf{u}}$ and $\hat{\mathbf{p}}$ are choices that define each
  panel, not fixed directions, and are indicated in each.
  (a) $p$-wave, $\hat{\mathbf{u}}=\hat{\mathbf{z}}$, $P_\delta=\sigma_0$: odd
  longitudinal field, $h_z\propto\sin k_x-\sin k_y$.
  (b) Altermagnet, $\hat{\mathbf{p}}=\hat{\mathbf{z}}$, $U_\delta=\sigma_0$: even
  longitudinal field of $d$-wave form, $h_z\propto\cos k_x-\cos k_y$.
  (c) Rashba, $\hat{\mathbf{u}}_x=\hat{\mathbf{y}}$, $\hat{\mathbf{u}}_y=-\hat{\mathbf{x}}$:
  odd in-plane winding, generated by the exchange alone with no relativistic
  spin--orbit coupling.   In (a)-(c) the vanishing component is identically zero.
  (d) Mixed: both factors act, with non-parallel axes $\hat{\mathbf{u}}=\hat{\mathbf{y}}$,
  $\hat{\mathbf{p}}=\hat{\mathbf{z}}$, so $\hat{\mathbf{u}}\times\hat{\mathbf{p}}=\hat{\mathbf{x}}$.
  Parallel axes would commute and give no mixed term; their non-collinearity is what
  makes $h_z$ and $\mathbf{h}_{\perp}$ finite together, producing the
  mixed-parity, noncoplanar spin-bond texture.}
\end{figure}

These familiar non-relativistic spin textures arise
from the bond pattern through the single field $\mathbf{h}(\mathbf{k})$.
Representative momentum-space textures are shown in Fig.~\ref{fig:spin_textures}
for the pure $p$-wave and altermagnetic limits, for the Rashba exchange texture,
and the mixed noncoplanar sector, whose
microscopic origin is discussed below. Panels (a) to (c) are confined to a single subspace of spin: purely longitudinal along the N\'eel axis in the two polar limits, purely transverse in the Rashba case. The mixed texture is the only one displaying longitudinal and transverse
spin-bond components simultaneously, see Fig.~\ref{fig:spin_textures} panel (d).

The vanishing entries in Fig.~\ref{fig:spin_textures} are structural. For the case with $\hat{\mathbf{u}}$ or $\hat{\mathbf{p}}$ along $\hat{\mathbf{z}}$,
the transverse component does not exist in the bond algebra, whereas for the in-plane
unitary axes the longitudinal one does not. Only when both polar sectors are
present and fail to commute do the two survive together. These textures are
therefore not isolated points but limits of a continuous family: changing the bond
axes carries one class into another, and moving continuously between an even and an
odd limit with non-parallel axes necessarily passes through the mixed sector. With
the construction validated against the known classes, we now map those crossovers
and then turn to the non-coplanar regime that has no counterpart among them.

\subsection*{Crossovers between magnetic classes}
The classes of Table~\ref{tab:recipes} are not isolated: a continuous change of the bond axes carries one into another. The orientation of the bond axis $\uhat$ relative to the N\'eel director $\Nhat$ is a continuous knob.

We can systematically map the transition from $p$-wave to Rashba physics by canting the bond axis away from the N\'eel vector. We introduce a canting angle $\theta$ towards an in-plane direction $\hat{\mathbf w}_\delta$,
\begin{equation}
\uhat_\delta(\theta)=\cos\theta\,\zhat+\sin\theta\,\hat{\mathbf w}_\delta ,
\end{equation}
where $\theta$ is a continuous knob: at $\theta=0$ the axis lies along the easy
axis $\zhat$ ($p$-wave), and at $\theta=90^\circ$ it lies fully in the plane
(Rashba-like). Physically, $\theta$ is set by N\'eel-vector reorientation,
through an applied magnetic field, strain, or canting of the local moments,
and is therefore experimentally accessible. The
spin feels a tug-of-war between two fields [Eq.~\eqref{eq:texture}]: the exchange
$\Delta\zhat$ pins it along the easy axis, while the bond field $\hh$ pulls it toward $\uhat$. At $\theta=0$ both point along $\zhat$ and the spin is collinear, a $p$-wave magnet; as $\theta$ grows, $\hh$ acquires an in-plane part $\propto\sin\theta$ and the spin tilts into the plane, winding with $\kk$ an emergent Rashba texture. The exact field in Eq.~\eqref{eq:texture} gives a polarization tilted from the easy axis by
\begin{equation}
\theta_{\rm spin}(\kk)=\arctan\frac{|\hh_\perp(\kk)|}{|\Delta-s\,h_z(\kk)|}\;\simeq\; \arctan\!\Big(\tfrac{t\sin\alpha}{\Delta}\sin\theta\Big),
\label{eq:tilt}
\end{equation}
in agreement with diagonalization ($\Delta/t=0.5\Rightarrow51^\circ$, $\Delta /t=3\Rightarrow11.6^\circ$). The crossover is a controllable interpolation
\begin{equation}
p\text{-wave magnet}\ \xrightarrow{\ \theta\ }\
\text{Rashba-like exchange spin texture},
\end{equation}
with a momentum-odd spin-ARPES signature $\langle s_\perp(\kk)\rangle\propto(t\sin\alpha/\Delta)\sin\theta$, sign-changing across $\Gamma$. N\'eel-vector reorientation of exactly this kind is now experimentally routine. Recent advances include nanoscale imaging and control in MnTe~\cite{Krempasky2024}, as well as electrical $180^\circ$ switching, spin--orbit and magnetic-octupole torques, and crystal-symmetry control in CrSb~\cite{Amin2024,Han2024,Zhou2025,Reimers2024}. Thus, the crossover knob is physically accessible. Throughout, it is the bond axis that is canted while the exchange remains collinear, so the exact solution still applies; canting of the moments themselves is a separate extension (Supplementary Note G).

The polar decomposition makes it straightforward to define a generalized Hamiltonian that continuously interpolates between all these classes:
\begin{enumerate}
\item \emph{$p$-wave $\to$ Rashba} and \emph{$p$-wave $\to$ Dresselhaus}: cant the
unitary axis from $\zhat$ toward the in-plane Rashba
($\uhat_x{=}\hat{\mathbf y},\uhat_y{=}{-}\hat{\mathbf x}$) or Dresselhaus
($\uhat_x{=}\hat{\mathbf x},\uhat_y{=}{-}\hat{\mathbf y}$) configuration, by angle
$\theta$; the spin tilts by $\arctan(t\sin\alpha\sin\theta/\Delta)$ in both cases
[Eq.~\eqref{eq:tilt}].
\item \emph{Rashba $\leftrightarrow$ Dresselhaus}: rotate the in-plane axes in
\emph{opposite} senses, $\uhat_x(\chi)=(-\sin\chi,-\cos\chi,0)$,
$\uhat_y(\chi)=(\cos\chi,\sin\chi,0)$; $\chi{=}0$ is Rashba, $\chi{=}90^\circ$ is
Dresselhaus, and $\chi{=}45^\circ$ is the equal Rashba--Dresselhaus point, a
persistent spin helix.
\item \emph{$d$-altermagnet $\leftrightarrow$ $p$-wave}: vary the weight between
$P$ and $U$ along the common axis $\zhat$; the operators commute, so the texture
stays collinear and the splitting evolves continuously from even ($d$) to odd
($p$).
\item \emph{$d$-altermagnet $\leftrightarrow$ Rashba}: vary the weight with
$\phat=\zhat$ (even) and the unitary axes in-plane (odd). Here
$[U,P]\neq0$, and the interpolation passes through a mixed, non-coplanar sector:
a transverse, even-in-momentum spin component along $\uhat\times\phat$ that is
nonzero only in the interior of the
crossover and vanishes at both pure endpoints.\end{enumerate}
This completes the unified picture of the collinear (commuting) phase and amplitude bonds. We now transition to the most general case, the mixed, non-commuting
sector developed below. The first three interpolations keep the spin axes parallel and stay within the commuting family of Table~\ref{tab:recipes}. The fourth does not: whenever the interpolation is carried out with non-parallel spin axes, the non-commuting mixed sector appears as the bridge between the even (altermagnetic) and odd ($p$-wave/SOC) limits. We develop that sector next.

\subsection*{Scope of the classification: altermagnetism and beyond}

We can now place the spin-bond classification precisely relative to altermagnetism. Altermagnetism, as defined by spin-group symmetry, is the even-parity collinear class~\cite{Smejkal2022a,Smejkal2022b}. In the present language, altermagnetism is a bond-structured realization of the Hermitian (spin-amplitude) sector,
\begin{equation}
U_\delta=\sigma_0,\ P_\delta\neq\sigma_0\ \Rightarrow\ H_{\rm alt}(\kk)=\lambda_d(\cos k_x-\cos k_y)\sigma_z,\quad H_{\rm alt}(-\kk)=H_{\rm alt}(\kk).
\end{equation}
The complementary unitary (spin-phase) sector, $U_\delta\neq\sigma_0,\ P_\delta=\sigma_0$, gives the odd-parity exchange spin textures with the symmetry of $p$-wave magnets~\cite{Hellenes2024}. When both sectors are present and \emph{commute}, the result is a coplanar superposition of an even and an odd part. When they fail to commute, $[U_\delta,P_{\delta}]\neq0$, the bond cannot be diagonalized along a fixed spin axis and the texture becomes \emph{non-coplanar}, a regime unreachable from any commuting (coplanar) bond set. The hierarchy is

\begin{equation}
\boxed{
\begin{aligned}
P_\delta = P,\qquad U_\delta = \sigma_0
&\Rightarrow \Gamma\text{-split Hermitian sector},\\
P_\delta\ \text{bond structured},\qquad U_\delta = \sigma_0
&\Rightarrow \text{even-parity altermagnet},\\
P_\delta = \sigma_0,\qquad U_\delta \neq \sigma_0
&\Rightarrow \text{odd-parity }p\text{-wave / exchange-SOC},\\
U_\delta,P_\delta \neq \sigma_0,\qquad [U_\delta,P_\delta]\neq 0
&\Rightarrow \text{mixed-parity, non-coplanar spin-bond magnet}.
\end{aligned}
}
\end{equation}

\
In this view, altermagnetism is not overturned, but rather recognized as the even-parity limit of a broader algebraic framework. This aligns with recent studies showing that the landscape of compensated spin-split magnets extends well beyond standard altermagnets. Symmetry analyses have identified spin-split antiferromagnets that fall outside this class~\cite{Yuan2024}, odd-parity magnetism has emerged as a distinct non-relativistic mechanism~\cite{Hellenes2024,Chakraborty2025}, and mixed-parity states have been proposed to bridge the two regimes~\cite{Zhuang2026}. The spin-bond polar decomposition supplies a microscopic, exactly solvable algebraic language for precisely this space: the unitary/Hermitian split fixes the parity, and the non-commutativity of the link operators is the diagnostic for the mixed, non-coplanar sector.

The most intricate physical phenomena emerge from this mixed regime, and its origin can be written down explicitly by expanding the bond operator to leading order in the spin couplings and using $[\uhat\!\cdot\!\bsig,\phat\!\cdot\!\bsig]=2i(\uhat\times\phat)\!\cdot\!\bsig$, we find:
\begin{equation}
T_\delta\simeq\sigma_0+i\alpha_\delta\,\uhat_\delta\!\cdot\!\bsig +\beta_\delta\,\phat_\delta\!\cdot\!\bsig \;\underbrace{-\,\alpha_\delta\beta_\delta\,(\uhat_\delta\times\phat_\delta)\!\cdot\ \bsig}_{\text{mixed term}}+\dots
\label{eq:mixterm}
\end{equation}

The first spin term is the unitary ($p$-wave) part; the second is the Hermitian (altermagnetic) part; the third, proportional to $\uhat_\delta\times\phat_\delta$, is present only when the two sectors fail to commute, and is absent in both a pure altermagnet ($U_\delta=\sigma_0$) and a pure $p$-wave magnet ($P_\delta=\sigma_0$). Its momentum-space fingerprint is an \emph{even-parity} spin component along $\uhat\times\phat$, transverse to the collinear axis, which vanishes identically when $[U_\delta,P_{\delta}]=0$. Combined with the longitudinal odd ($p$-wave) and even (altermagnetic) parts, this transverse term tilts the texture out of a single plane: it is the microscopic seed of the non-coplanar, mixed-parity regime.

To capture this non-coplanar mixing systematically, we introduce the spin-bond generator $\mathcal G_\delta=\log T_\delta$, which in Pauli form reads
\begin{equation}
\mathcal G_\delta=\big(\mathbf b_\delta+i\,\mathbf a_\delta\big)\cdot\boldsymbol\sigma ,
\label{eq:gen_ab}
\end{equation}
the real vector $\mathbf a_\delta$ is the exchange-generated spin-phase field of the bond (its unitary, $p$-wave content) and $\mathbf b_\delta$ is its spin-amplitude field (its Hermitian, altermagnetic content). In the commuting limit the two are independent. When the bond axes do not commute they mix, and the conversion goes both ways: non-commutativity turns spin phase into spin amplitude, and spin amplitude into spin phase.

The leading effect is the cross term
\begin{equation}
\mathcal G_\delta^{\rm amp}
=
-\alpha_\delta\beta_\delta\,
(\hat{\mathbf u}_\delta\times\hat{\mathbf p}_\delta)\cdot\boldsymbol\sigma ,
\label{eq:gen_amp}
\end{equation}
which is \emph{Hermitian}: a phase and an amplitude combine to produce a new spin
\emph{amplitude} along the emergent axis
$\hat{\mathbf u}_\delta\times\hat{\mathbf p}_\delta$. The reciprocal conversion
appears one order higher and is \emph{anti-Hermitian}, i.e.\ phase-like,
\begin{equation}
\mathcal G_\delta^{\rm ph} =\frac{i\,\alpha_\delta\beta_\delta^{2}}{3}\,
\hat{\mathbf u}^{(p)}_{\delta,\perp}\cdot\boldsymbol\sigma ,
\qquad \hat{\mathbf u}^{(p)}_{\delta,\perp} =
\hat{\mathbf u}_\delta-(\hat{\mathbf u}_\delta\cdot\hat{\mathbf p} _\delta)\, \hat{\mathbf p}_\delta ,
\label{eq:gen_ph}
\end{equation}
a new spin \emph{phase} along the component of $\hat{\mathbf u}_\delta$ transverse to $\hat{\mathbf p}_\delta$. The phase-to-amplitude channel in Eq.~\eqref{eq:gen_amp} is the leading one; the reciprocal amplitude-to-phase channel in Eq.~\eqref{eq:gen_ph} is its higher-order counterpart. The mixed sector is therefore not a sum of a unitary and a Hermitian link, but a phase--amplitude conversion produced by the non-commuting algebra of the bond. The local strength of the mixing is measured by a single invariant of the bond,
\begin{equation}
\chi_{{\rm mix},\delta}=\big|\mathbf a_\delta\times\mathbf b_\delta\big| \simeq\alpha_\delta\beta_\delta\, \big|\hat{\mathbf u}_\delta\times\hat{\mathbf p}_\delta\big| ,
\label{eq:chi_mix_body}
\end{equation}

which vanishes when the spin-phase and spin-amplitude axes are parallel and is
largest when they are orthogonal. The full generator expansion, including the higher-order back-action terms, is given in Supplementary Note H. In addition, a relativistic spin--orbit contribution enters the same
unitary sector and, for fixed $P_{\delta}$, renormalizes and rotates
the total spin-phase vector $\mathbf{a}_{\delta}^{\mathrm{tot}}$. It
therefore tunes the mixing invariant
$\chi_{\mathrm{mix},\delta}^{\mathrm{tot}}
=
|\mathbf{a}_{\delta}^{\mathrm{tot}}\times\mathbf{b}_{\delta}|$,
providing an additional means to activate or suppress the mixed sector. Taken together, these results show that the significance of the polar
decomposition goes beyond producing an additional spin texture. It
provides an organizing principle for compensated spin-split magnets:
the Hermitian and unitary factors identify the even- and odd-parity
limits, while their non-commutativity identifies the mixed sector.
Spin textures that might otherwise appear as unrelated experimental
observations are thus placed within a single bond-level classification,
with a clear criterion for when the mixed component is forbidden,
allowed, or required along a non-collinear crossover.

\begin{figure*}[tbh]
  \centering
  \includegraphics[width=\textwidth]{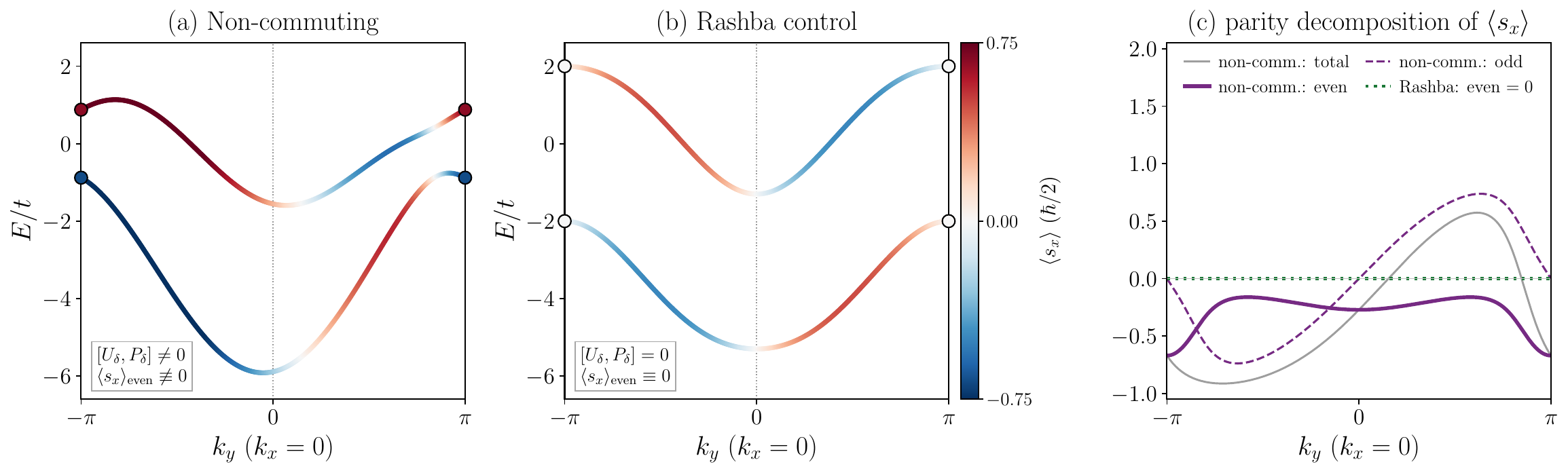}
  \caption{\label{fig:arpes}
  Spin-ARPES fingerprint of the non-commuting spin-bond sector. Bands along
$k_x=0$, coloured by the transverse polarization $\langle s_\perp\rangle$ projected
onto $\hat{\mathbf{u}}\times\hat{\mathbf{p}}=\hat{\mathbf{x}}$, within the
sublattice-parity sector $s=+1$. Parameters are $\alpha=0.6$, $\beta=0.5$, and
$\Delta/t=2$; momenta are in units of the inverse lattice constant $a=1$, energies in
units of $t$, and spin polarization in units of $\hbar/2$. Circles mark the equivalent
zone-boundary points $k_y=\pm\pi$. (a) In the mixed non-commuting sector, the lower band
has the same-sign polarization at both markers, and the bands are not symmetric about
$k_y=0$. (b) In the Rashba control, the polarization is odd in momentum and vanishes
at the zone boundary, so the markers are colourless. (c) Even- and odd-in-momentum
components of the lower-band polarization, from
Eq.~\eqref{eq:parity_decomposition}. The even component is finite only in the mixed
sector and vanishes identically for Rashba.
}
\end{figure*}

A direct experimental consequence of this classification is a spin-ARPES signature. The relevant observable is the band spin projected onto
$\hat{\mathbf{u}}\times\hat{\mathbf{p}}$ which is the axis perpendicular to both the unitary
($p$-wave) and Hermitian (altermagnetic) bond axes. We denote this projection by the
transverse polarization $\langle s_\perp\rangle(\mathbf{k})$ and decompose it into
momentum-even and momentum-odd components,
\begin{equation}
\langle s_\perp\rangle_{\mathrm{even/odd}}(\mathbf{k})
=
\frac{1}{2}
\left[
\langle s_\perp\rangle(\mathbf{k})
\pm
\langle s_\perp\rangle(-\mathbf{k})
\right],
\label{eq:parity_decomposition}
\end{equation}
as shown in Fig.~\ref{fig:arpes}(c) (see Methods). For the minimal bond pattern
considered here, the mixed-sector criterion is
\begin{equation}
\langle s_\perp\rangle_{\mathrm{even}}(\mathbf{k})
\not\equiv 0
\quad\Longleftrightarrow\quad
[U_\delta,P_{\delta}]\neq0 .
\label{eq:fingerprint}
\end{equation}

What makes $\langle s_\perp\rangle_{\mathrm{even}}$ so relevant, is the simultaneous
combination of transverse spin polarization and even momentum parity. Within the pure
limits of the present classification, an altermagnet is even but longitudinal, a
$p$-wave magnet is odd and longitudinal, and Rashba or Dresselhaus textures are odd
and transverse. The simultaneous occurrence of both attributes therefore identifies
the non-commuting mixed sector. Its magnitude grows with
$|\hat{\mathbf{u}}_\delta\times\hat{\mathbf{p}}_\delta|$ and, to leading order in the
couplings, is proportional to $\chi_{\mathrm{mix},\delta}$,
vanishing continuously as the axes become parallel and $[U,P]\rightarrow0$.

For the illustrative bond couplings of Fig.~\ref{fig:arpes}
($\alpha=0.6$, $\beta=0.5$), the peak even-transverse signal is
sizeable, reaching approximately $0.80$ in units of $\hbar/2$ near
$\Delta/t\simeq1.6$ and $0.67$ at $\Delta/t=2$. For larger exchange,
it crosses over to a decay proportional to $t\alpha\beta/\Delta$,
reaching approximately $0.09$ at $\Delta/t=8$. The size of the signal follows from a simple ratio. The non-commuting bonds produce a
transverse field of strength $2t\alpha\beta$, which competes with the exchange
$\Delta$ that holds the spin along the N\'eel axis. When the exchange dominates, the
spin is tilted from that axis by the ratio of the two,
\begin{equation}
\max_{\mathbf{k}}\left|\langle s_\perp\rangle_{\mathrm{even}}\right|
\simeq
\frac{2t\alpha\beta}{\Delta},
\label{eq:fingerprint_scaling}
\end{equation}
so a larger spin splitting or weaker exchange gives a larger fingerprint.

The signal is largest near the Brillouin-zone boundary. Along $k_x=0$, the points
$k_y=\pm\pi$ are invariant under momentum inversion modulo a reciprocal-lattice
vector. Consequently, $\langle s_\perp\rangle_{\mathrm{odd}}$ vanishes there, and the
measured transverse polarization directly equals its even component
[Fig.~\ref{fig:arpes}(c)]. Thus, with the detection axis chosen along
$\hat{\mathbf{u}}\times\hat{\mathbf{p}}$ and the spin-split bands resolved, a nonzero
even-in-momentum transverse polarization is the direct fingerprint of the non-commuting
sector. In the minimal model, the mixed texture also produces a secondary, non-unique
signature: within a fixed sublattice-parity sector, the dispersion can become
asymmetric under $\mathbf{k}\rightarrow-\mathbf{k}$ [Fig.~\ref{fig:arpes}(a)].

\section*{Discussion}
We have introduced a spin-bond theory of non-relativistic spin
splitting, where the central microscopic object is the bond operator
$T_\delta$. The polar decomposition
\begin{equation}
    T_\delta=U_\delta P_\delta
\end{equation}
provides the organizing principle. The unitary factor \(U_\delta\) carries
spin-dependent phases and generates odd-in-momentum spin fields, corresponding to \(p\)-wave exchange textures and emergent non-relativistic spin--orbit coupling. The Hermitian positive factor \(P_\delta\) carries spin-dependent amplitudes and generates even-in-momentum spin fields, containing altermagnetism as its natural limit. Thus \(p\)-wave magnetism, exchange-SOC textures, and altermagnetism appear as different algebraic limits of the same spin-bond object. For reciprocal bipartite lattices, the Hamiltonian factorizes exactly into sublattice-parity sectors. In each sector, the spin texture is controlled by the
effective field
\begin{equation}
    \Delta\hat{\mathbf z}-s\,\mathbf h(\mathbf k),
\end{equation}
demonstrating that the observed spin splitting is determined by the competition between the uniform exchange field and the momentum-dependent spin-bond field. This makes the parity locking transparent: unitary links produce sine form factors and odd spin textures, whereas Hermitian links produce cosine form factors and even spin
textures.
The framework also identifies a mixed regime beyond the pure limits. When
\(U_\delta\) and \(P_\delta\) do not commute, the bond contains the leading mixed
term
\begin{equation}
    -\alpha_\delta\beta_\delta \left( \hat{\mathbf u}_\delta \times \hat{\mathbf p}_\delta \right)\cdot\boldsymbol\sigma,
\end{equation}
which is absent in both pure altermagnets and pure \(p\)-wave magnets. This term
creates a transverse spin axis and is the microscopic seed of a mixed-parity,
non-coplanar spin-bond texture. Equivalently, the generator
\(\mathcal G_\delta=\log T_\delta\) reveals this regime as a phase--amplitude
conversion between the unitary and Hermitian sectors, with a natural mixing scale
\begin{equation}
    \chi_{{\rm mix},\delta}=\left|\mathbf a_\delta\times\mathbf b_\delta\right|\simeq\alpha_\delta\beta_\delta\left|\hat{\mathbf u}_\delta\times \hat{\mathbf p}_\delta \right| .
\end{equation}

The most direct consequence of the non commuting sector is a spectroscopic one. In the minimal model, it produces an even-in-momentum transverse spin polarization,
\begin{equation}
    \langle s_\perp\rangle_{\rm even}(\mathbf k)\neq0,
\end{equation}
along the axis
\(\hat{\mathbf u}\times\hat{\mathbf p}\). This component is absent in the pure altermagnetic, pure \(p\)-wave, Rashba, and Dresselhaus limits, and therefore provides a falsifiable spin-ARPES fingerprint of non-commuting spin-bond
magnetism.

These results establish the spin-bond operator $T_{\delta}$ as a
constructive principle for non-relativistic spin splitting. Its polar
factors fix the parity and geometry of the spin texture, and their
non-commutation generates a mixed regime with a transverse,
even-in-momentum polarization. What makes this a principle rather than
a model is that $T_{\delta}$ is the physical input of the theory: it
may arise from orbital downfolding, inequivalent hopping paths,
non-collinear moments, or interfacial environments, and the resulting
classification is fixed by its algebra alone. Understanding where a
texture comes from becomes a way of specifying one: choosing
the polar content of the links and the relative orientation of their
spin axes selects the parity, the spin geometry, and the degree of
non-coplanarity, opening a route to non-relativistic spin textures that
do not correspond to any of the established classes so far.

Beyond the fundamental classification of magnetic phases, the spin-bond formalism connects naturally to spin-based quantum computing, offering a useful microscopic language for the emerging field of altermagnet-based qubit architectures \cite{AbadilloUriel2026,Kirczenow2026,Vakili2026,VosoughiNia2025,Steinacker2025}. The discovery of the non-commuting mixed-parity regime ($[U_\delta, P_{\delta}] \neq 0$) suggests that synthetic spin-orbit coupling can be deterministically induced or suppressed via geometric tuning, see Supplementary Note I. Current proposals for gate-defined spin qubits \cite{Burkard2023,Vandersypen2017,Hetenyi2020} rely on micromagnet field gradients \cite{PioroLadriere2008} or persistent spin-orbit fields to perform all-electrical qubit control via Electric Dipole Spin Resonance (EDSR)~\cite{Golovach2006,Nowack2007,NadjPerge2010,AbadilloUriel2026,Vakili2026}. Our theory offers a distinct microscopic alternative: by dynamically tuning the bond sectors, the synthetic spin-orbit coupling can be switched ``on-demand''. This enables a protocol where the spin-orbit field is turned on exclusively during EDSR gating for ultra-fast operations, and subsequently turned off ($[U_\delta, P_{\delta}] \to 0$) to recover a purely collinear, decoherence-free quantum memory~\cite{Sun2026}, see Supplementary Note J. While predicting the precise EDSR Rabi frequencies requires material-specific \textit{ab initio} calculations, the bond-operator formalism guarantees that the field is fundamentally tuneable. This provides a concrete lattice-level blueprint for engineering highly coherent, field-free spin qubits.

\section*{Methods}

\paragraph{Model and parameters.}
In our tight-binding implementation, we define the unitary and Hermitian bond operators using a general exponential parameterization. Specifically, the unitary part carrying spin-dependent phases is written as $U(\hat{\mathbf n},a)=e^{ia\,\hat{\mathbf n}\cdot\bsig} =\cos a+i\sin a\,(\hat{\mathbf n}\cdot\bsig)$, while the Hermitian part carrying spin-dependent amplitudes is defined as $P(\hat{\mathbf n},b)=e^{b\,\hat{\mathbf n}\cdot\bsig}=\cosh b+\sinh b\,(\hat{\mathbf n}\cdot\bsig)$. To study the mixed non-commuting sector, we set the unitary ($p$-wave) spin axis to $\uhat=\hat{\mathbf y}$ and the Hermitian (altermagnetic) spin axis to $\phat=\zhat$. The emergent transverse direction is therefore given by the cross product $\uhat\times\phat=\hat{\mathbf x}$. For a representative square lattice model, the bond operators along the $\hat{\mathbf x}$ and $\hat{\mathbf y}$ directions are chosen as:
\begin{equation}
T_x=U(\hat{\mathbf y},\alpha)\,P(\zhat,\beta),\qquad
T_y=U(-\hat{\mathbf x},\alpha)\,P(\zhat,-\beta),
\end{equation}
where we use dimensionless coupling strengths $\alpha=0.6$ and $\beta=0.5$. The scalar hopping is set to $t=1$, and the local exchange field is chosen to be $\Delta=2$. The corresponding momentum-space Bloch hopping matrix is

\begin{equation}
S(\kk)=T_x e^{ik_x}+T_x^\dagger e^{-ik_x}+T_y e^{ik_y}+T_y^\dagger e^{-ik_y}.
\end{equation}
From this matrix, we extract the effective scalar field $h_0(\kk)=\tfrac12\mathrm{Tr}[S(\kk)]$ and the corresponding spin-vector field $\hh(\kk)=\tfrac12\mathrm{Tr}[S(\kk)\boldsymbol\sigma]$.

\paragraph{Sector-resolved spin texture.}
A critical aspect of extracting the correct spin polarization is the resolution of the sublattice-parity sectors, labeled by $s=\pm 1$. Rather than simply computing the expectation value for the globally lowest band of the $4\times4$ Hamiltonian, we strictly evaluate the texture within a fixed parity sector. Failing to do so introduces a band-crossing artifact across the Brillouin zone, which scrambles the parity decomposition and yields a spurious even-transverse spin signal even in purely odd-parity systems (like the Rashba limit). By fixing the sector (e.g., $s=+1$), the lower-band polarization can be obtained analytically as derived in the Supplementary Information:
\begin{equation}
\langle\bm s\rangle_{s}(\kk)=-\frac{\Delta\zhat-s\,\hh(\kk)}{|\Delta\zhat-s\,\hh(\kk)|},
\end{equation}
evaluated here at $s=+1$. Experimentally, this sector-specific resolution corresponds exactly to the process of measuring individual spin-split bands using spin- and angle-resolved photoemission spectroscopy (spin-ARPES).

\paragraph{Observable and decomposition.}
To isolate the non-coplanar signature of the mixed regime, we project the computed spin polarization onto the emergent transverse axis, $\uhat\times\phat=\hat{\mathbf x}$. This projected component, $\langle s_x\rangle(\kk)$, is then explicitly decomposed into its momentum-even and momentum-odd parts:
\begin{equation}
\langle s_x\rangle_{\rm even/odd}(\kk)=\tfrac12\big[\langle s_x\rangle(\kk) \pm\langle s_x\rangle(-\kk)\big],
\end{equation}
This numerical decomposition is performed over a uniform $k$-grid spanning the
Brillouin zone. Results are converged with respect to the mesh; the values reported
below are identical on grids from $60\times60$ to $960\times960$. The primary observable reported in our analysis is the maximum amplitude of the even-parity transverse polarization, defined as $\max_{\kk}|\langle s_x\rangle_{\rm even}|$.

\paragraph{Verification against control models.}
To confirm the uniqueness of the non-commuting fingerprint, we applied the identical decomposition procedure to all the pure, commuting magnetic classes. For a pure altermagnet ($T_x=P(\zhat,\beta), T_y=P(\zhat,-\beta)$), a pure $p$-wave magnet ($T_x=U(\zhat,\alpha),T_y=U(\zhat,-\alpha)$), and a pure Rashba system ($T_x=U(\hat{\mathbf y},\alpha),T_y=U(-\hat{\mathbf x},\alpha)$), the even-transverse polarization vanishes.
Conversely, the mixed non-commuting sector yields a non-zero value ($\approx0.67$ here) for the parameters specified above. This confirms the fundamental relation $\langle s_\perp\rangle_{\rm even}\neq0\Leftrightarrow[U_\delta,P_{\delta}]\neq0$.

\paragraph{Scaling, spatial distribution, and tunability}
We systematically mapped the behavior of the even-transverse polarization across different parameter regimes.  The peak signal is 0.54 at $\Delta/t=1$, rises to a maximum of $\approx0.80$ near
$\Delta/t\simeq1.6$, and is 0.67 at $\Delta/t=2$. For stronger exchange it decays as
$t\sin\alpha/\Delta$, taking 0.39, 0.24, 0.14 and 0.093 at $\Delta/t=3,4,6$ and 8.  In momentum space, this peak polarization is primarily concentrated along the Brillouin-zone boundaries, specifically near the $\kk=(0,\pm\pi)$ points for our chosen axes. Most importantly, the signature can be deterministically switched off: if the respective spin axes are brought into parallel alignment ($\phat\to\uhat$, ensuring $[U,P]\to0$), the transverse signal continuously vanishes. Although the precise decay profile depends on the reshaping of the overall texture, this behavior provides a robust mechanism for the on/off control of the synthetic spin-orbit coupling via geometric tuning.

\section*{Acknowledgements}
S.A. acknowledges funding from Fondecyt Regular 1261323 and ANID CEDENNA CIA 250002. This work is dedicated to the memory of Dominique Givord and Konstantin Y. Gusliyenko, whose encouragement and scientific vision remain with us.

\renewcommand{\thefigure}{S\arabic{figure}}
\renewcommand{\theequation}{S\arabic{equation}}
\renewcommand{\thesection}{S\arabic{section}}
\setcounter{figure}{0}
\setcounter{equation}{0}

\clearpage
\appendix

\begin{center}
    {\Large\bfseries APPENDIX}
\end{center}

\vspace{0.5cm}

\section{Derivation of the Hamiltonian}

\label{app:fourier}

Let $c_{A\sigma}$ and $c_{B\sigma}$ be the electron annihilation operators with spin $\sigma$ in momentum space on sublattices $A$ and $B$. Then we can write the operators $c_{i\sigma}$ and $c_{j\sigma}$ as:

\begin{equation}
c_{i\sigma} = \frac{1}{\sqrt{N_A}}
\sum_{\mathbf{k}} e^{+i\mathbf{k}\cdot \mathbf{R}_i}\, c_{A\sigma}(\mathbf{k}) \quad (i\in A)
\end{equation}

\begin{equation}
c_{j\sigma} = \frac{1}{\sqrt{N_B}}
\sum_{\mathbf{k}} e^{+i\mathbf{k}\cdot \mathbf{R}_j}\, c_{B\sigma}(\mathbf{k}) \quad (j\in B)
\end{equation} 

Then the first term of Eq. 4 of the main text, or hopping term, takes the form:

\begin{align}
H_{\mathrm{hop}}&=
- \sum_{\langle i\in A, j\in B\rangle,\sigma}
\Big( t_{ij}\, c_{i\sigma}^\dagger c_{j\sigma} + \text{h.c.}\Big)\\
&= - \sum_{\langle i\in A, j\in B\rangle,\sigma}
\frac{t_{ij}}{\sqrt{N_A N_B}}
\sum_{\mathbf{k},\mathbf{k}'}
e^{-i\mathbf{k}\cdot \mathbf{R}_i}\, e^{+i\mathbf{k}'\cdot \mathbf{R}_j}\,
c_{A\sigma}^\dagger(\mathbf{k}) c_{B\sigma}(\mathbf{k}')
+\text{h.c.}
\\
&= - \sum_{\sigma}\frac{1}{\sqrt{N_A N_B}}
\sum_{i\in A}\sum_{j\in \mathrm{NN}(i)\cap B} 
\sum_{\mathbf{k},\mathbf{k}'}
t_{ij}\, e^{i(\mathbf{k}'-\mathbf{k})\cdot \mathbf{R}_i}\,
e^{+i\mathbf{k}'\cdot (\mathbf{R}_j-\mathbf{R}_i)}\,
c_{A\sigma}^\dagger(\mathbf{k}) c_{B\sigma}(\mathbf{k}')
+\text{h.c.}
\end{align}

In a periodic lattice, the hopping amplitude only depends on the relative distance $\delta = \mathbf{R}_j-\mathbf{R}_i$, so we can set $t_{ij}=t_\delta$. This allows us to write $\sum_{j\in \mathrm{NN}(i)\cap B} t_{ij} e^{+i\mathbf{k}'\cdot (\mathbf{R}_j-\mathbf{R}_i)} =\sum_{\delta} t_\delta e^{+i\mathbf{k}'\cdot \delta}$. Furthermore, we can use $\sum_{i\in A} e^{i(\mathbf{k}-\mathbf{k}')\cdot \mathbf{R}_i} 
= N_A\,\delta_{\mathbf{k},\mathbf{k}'}$. Therefore, we have:

\begin{equation}
H_{\mathrm{hop}}=
- \sum_{\sigma}\sum_{\mathbf{k}}
g(\mathbf{k})\,
c_{A\sigma}^\dagger(\mathbf{k}) c_{B\sigma}(\mathbf{k})
+\text{h.c.},
\end{equation}
where we have assumed $N_A=N_B$ and defined the structure factor:

\begin{equation}
g(\mathbf{k}) \;=\; \sum_{\delta} t_{\delta}\, e^{+i\mathbf{k}\cdot \delta}.
\end{equation}

Let us define the spinor as $Psi_{\mathbf{k}\sigma} =
\begin{pmatrix}
c_{A\sigma}(\mathbf{k}) \\
c_{B\sigma}(\mathbf{k})
\end{pmatrix}$. Then:

\begin{equation}
H_{\text{hop}}
= -\sum_{\mathbf{k},\sigma}
\Psi_{\mathbf{k}\sigma}^\dagger
\begin{pmatrix}
0 & g(\mathbf{k}) \\
g^*(\mathbf{k}) & 0
\end{pmatrix}
\Psi_{\mathbf{k}\sigma}.
\end{equation}

Let us now define $\tau_x$, $\tau_y$, and $\tau_z$ as the Pauli matrices in the sublattice space $(A,B)$. If we define $\tau_\pm = \frac{\tau_x \pm i \tau_y}{2}$ as the raising $\tau_+$ and lowering $\tau_-$ matrices in the $(A,B)$ space, then $H_{\text{hop}}(\mathbf{k})$ takes the form:

\begin{equation}
H_{\text{hop}}(\mathbf{k})
= -(\tau_+\otimes\sigma_0)\,g(\mathbf{k})
- (\tau_-\otimes\sigma_0)\,g^*(\mathbf{k}),
\end{equation}

where $\sigma_0$ is the identity in real spin space and we have used $c_{A\sigma}^\dagger c_{B\sigma}
= \Psi_{\mathbf{k}\sigma}^\dagger \,\tau_+\, \Psi_{\mathbf{k}\sigma}$ and $c_{B\sigma}^\dagger c_{A\sigma}
= \Psi_{\mathbf{k}\sigma}^\dagger \,\tau_-\, \Psi_{\mathbf{k}\sigma}$.
The last two terms of Eq. 4 of the main text remain. In this case we have:

\begin{align}
H_{\mathrm{ex}} &=  \sum_{i\in A} \Delta\, c_i^\dagger \sigma_z c_i
- \sum_{j\in B} \Delta\, c_j^\dagger \sigma_z c_j\\
&=+ \sum_{\mathbf{k}} \Delta\, c_A^\dagger(\mathbf{k})\, \sigma_z\, c_A(\mathbf{k})-\sum_{\mathbf{k}} \Delta\, c_B^\dagger(\mathbf{k})\, \sigma_z\, c_B(\mathbf{k}).
\end{align}

Thus $H_{\mathrm{ex}}(\mathbf{k}) = \Delta\,
 \big(\tau_z\otimes \sigma_z\big)$, where $\sigma_z$ is the $z$-component Pauli matrix in real spin space.

 Then the total Hamiltonian is

 \begin{equation}
H(\mathbf{k})=-(\tau_+\otimes\sigma_0)\,g(\mathbf{k})
- (\tau_-\otimes\sigma_0)\,g^*(\mathbf{k})+\Delta\,
 \big(\tau_z\otimes \sigma_z\big).
 \end{equation}



\section{Uniform versus bond-dependent links}
\label{app:links}

This appendix shows that a bond-independent unitary dressing generated solely by the spin-frame transformation is a removable gauge artefact. This statement does not extend to a common Hermitian factor, which changes the spin-dependent hopping amplitudes and can produce physical spin splitting.

\paragraph{Removable uniform dressing.}
Start from the scalar antiferromagnet of Eq. 4 of the main text and perform a single spin rotation on each sublattice,
\begin{equation}
U
=
P_A\otimes U_A
+
P_B\otimes U_B,
\qquad
P_{A,B}
=
\frac{\tau_0\pm\tau_z}{2},
\label{eq:sublattice_rotation}
\end{equation}
where the projectors \(P_A\) and \(P_B\) act as sublattice selectors
(\(P_A\) acts only on \(A\), \(P_B\) only on \(B\)), so that \(U_A\) rotates the
spin on every \(A\) site and \(U_B\) on every \(B\) site by the same amount.
Because the original hopping is spin independent, \(T_{ij}=t_{ij}\sigma_0\),
every inter-sublattice bond is dressed by one and the same matrix,
\begin{equation}
\widetilde T_{ij}
=
U_A^\dagger\,(t_{ij}\sigma_0)\,U_B
=
t_{ij}\,W ,
\qquad
W=U_A^\dagger U_B ,
\label{eq:uniform_W}
\end{equation}
independent of the bond direction or of the position in the crystal. In momentum
space this gives
\begin{equation}
S(\mathbf k)
=
\sum_\delta t_\delta\,W\,e^{i\mathbf k\cdot\delta}
=
W\,g(\mathbf k),
\end{equation}
with \(g(\mathbf k)\) the scalar form factor of Eq. 7 of the main text. A single matrix that multiplies every bond can be factored out and undone by the global change of basis in sublattice space. It therefore leaves the spectrum \(E_\pm(\mathbf k)=\pm\sqrt{|g(\mathbf k)|^2+\Delta^2}\)
unchanged, and produces no spin splitting and no emergent spin-orbit coupling. Because \(W=U_A^\dagger U_B\) is unitary and generated solely by the spin-frame transformation, this uniform \(W\) is a gauge choice rather than a physical interaction.

\paragraph{Non-removable bond dependence.}
Genuine effects require the dressing to depend on the bond. Suppose the
\(x\)-directed and \(y\)-directed links connect to nonequivalent magnetic or
orbital environments, so that
\begin{equation}
\widetilde T_x
=
U_A^\dagger U_{B_x},
\qquad
\widetilde T_y
=
U_A^\dagger U_{B_y},
\qquad
\widetilde T_x \neq \widetilde T_y.
\label{eq:bond_dependent_W}
\end{equation}
Now no single matrix multiplies every bond, so there is no global rotation that
removes the spin structure simultaneously from all links. The momentum-space
operator
\begin{equation}
S(\mathbf k)
=
\sum_\delta \widetilde T_\delta\,e^{i\mathbf k\cdot\delta}
\end{equation}
no longer factorizes as \(W\,g(\mathbf k)\), and its non-scalar part survives in
the spectrum as a genuine, momentum-dependent spin splitting. This is the bond
dependence that cannot be gauged away, and it is the physical input isolated in
the main text.

\section{Minimal microscopic realization of the spin-bond operator}

The polar factors of the spin-bond operator $T_\delta$ fix the parity and geometry of the spin texture. Here we
exhibit one mechanism that produces the polar form $T_\delta=t_\delta U_\delta P_\delta$ explicitly, so that the polar factors acquire an explicit microscopic realization rather than remaining purely phenomenological ingredients. This minimal model is intended to explain the microscopic origin of a single spin-dependent bond.

We consider an electron that moves from site $i$ to site $j=i+\delta$ through an intermediate orbital $| d_\delta\rangle$. This intermediate state lies at higher energy and is therefore only virtually occupied,
\begin{equation} i \;\longrightarrow\; |d_\delta\rangle \;\longrightarrow\; j=i+\delta . \end{equation}
The spin dynamics acquired during this virtual hopping process is encoded in the resulting effective bond and gives rise to the two polar components introduced below.
The intermediate orbital has an energy $\Delta_\delta$ measured with respect to the low-energy states. In addition, the magnetic environment produces an exchange field acting on the spin of an electron that virtually occupies this orbital. The Hamiltonian of the intermediate orbital is therefore
\begin{equation}
H_{d,\delta}
=
\bm d_\delta^\dagger
\left[
\Delta_\delta \sigma_0
-
J_\delta \hat{\bm p}_\delta\cdot\bm{\sigma}
\right]
\bm d_\delta ,
\label{eq:Hd}
\end{equation}
where $J_\delta$ measures the strength of the exchange interaction and $\hat{\bm p}_\delta$ gives the direction of the corresponding local magnetic field. The first term sets the energy cost for virtually
occupying the intermediate orbital, while the second term makes this energy dependent on the spin orientation. Here, we introduce the fermionic spinor
\begin{equation}
\bm d_\delta =
\begin{pmatrix}
d_{\delta\uparrow} \\
d_{\delta\downarrow}
\end{pmatrix},
\end{equation}
where $d_{\delta\sigma}$ annihilates an electron with spin $\sigma$ in the intermediate orbital associated with bond $\delta$, while $d_{\delta\sigma}^\dagger$ denotes the corresponding creation operator. The exchange interaction splits the intermediate state into two spin states with energies
\begin{equation}
E_{\pm}=\Delta_\delta\mp J_\delta .
\end{equation}
We assume $\Delta_\delta>|J_\delta|$, so that both states remain outside the low-energy sector and are only virtually occupied during the hopping process. 

\paragraph{The Hermitian factor.}
The electron can hop between a magnetic site and the intermediate orbital with hopping amplitude $v_\delta$. Consequently, the effective hopping between $i$ and $j$ is proportional to $v_\delta^2$. Then, an electron can virtually hop through the intermediate orbital with its spin parallel or antiparallel to $\hat{\bm p}_\delta$. 

We denote by $t_\pm$ the resulting effective hopping amplitudes between sites $i$ and $j$ for these two spin orientations. From second-order perturbation theory,
\begin{equation}
t_\pm
=
\frac{v_\delta^2}{E_\pm}
=
\frac{v_\delta^2}{\Delta_\delta\mp J_\delta}.
\label{eq:tpm}
\end{equation}
The virtual occupation of the intermediate orbital generates two different effective hopping amplitudes between sites $i$ and $j$: $t_+$ for a spin parallel to $\hat{\bm p}_\delta$ and $t_-$ for a spin antiparallel to $\hat{\bm p}_\delta$. We now combine these two effective hopping amplitudes into a single operator acting on the spin.
The two effective hopping amplitudes can be parametrized as
$t_{+}=t_{\delta}e^{\beta_{\delta}}$ and
$t_{-}=t_{\delta}e^{-\beta_{\delta}}$, where
\begin{equation}
t_{\delta}=\sqrt{t_{+}t_{-}},
\qquad
\beta_{\delta}
=
\frac{1}{2}\ln\left(\frac{t_{+}}{t_{-}}\right).
\end{equation}
Here, $t_{\delta}$ is the scalar part of the effective hopping (
acts trivially on the spin), while $\beta_{\delta}$ measures how much more easily one spin orientation
is transmitted than the other, $t_{\pm}=t_{\delta}e^{\pm\beta_{\delta}}$.

We now construct a single hopping operator that gives $t_{+}$ for a spin
parallel to $\hat{\bm p}_{\delta}$ and $t_{-}$ for a spin antiparallel
to it. Since $\hat{\bm p}_{\delta}\cdot\bm{\sigma}$ has eigenvalue $+1$ for the
parallel spin and $-1$ for the antiparallel spin, this operator is
\begin{equation}
T_{\delta}^{(P)}
=
\frac{t_{+}+t_{-}}{2}\,\sigma_{0}
+
\frac{t_{+}-t_{-}}{2}\,
\hat{\bm p}_{\delta}\cdot\bm{\sigma}.
\label{eq:TP}
\end{equation}
Indeed, for the parallel spin this expression gives
\begin{equation}
\frac{t_{+}+t_{-}}{2}
+
\frac{t_{+}-t_{-}}{2}
=
t_{+},
\end{equation}
whereas for the antiparallel spin it gives
\begin{equation}
\frac{t_{+}+t_{-}}{2}
-
\frac{t_{+}-t_{-}}{2}
=
t_{-}.
\end{equation}

Using $t_{\pm}=t_{\delta}e^{\pm\beta_{\delta}}$, Eq.~\eqref{eq:TP}
becomes
\begin{equation}
T_{\delta}^{(P)}
=
t_{\delta}
\left[
\cosh\beta_{\delta}\,\sigma_{0}
+
\sinh\beta_{\delta}\,
\hat{\bm p}_{\delta}\cdot\bm{\sigma}
\right].
\end{equation}
Since $(\hat{\bm p}_{\delta}\cdot\bm{\sigma})^{2}=\sigma_{0}$, the
expression in brackets is precisely
\begin{equation}
\exp\left(
\beta_{\delta}\hat{\bm p}_{\delta}\cdot\bm{\sigma}
\right).
\end{equation}
Therefore,
\begin{equation}
T_{\delta}^{(P)}
=
t_{\delta}P'_{\delta},
\qquad
P'_{\delta}
=
\exp\left(
\beta_{\delta}\hat{\bm p}_{\delta}\cdot\bm{\sigma}
\right).
\end{equation}

In this microscopic model, $\beta_{\delta}$ is determined by the asymmetry between the two effective hopping amplitudes, $\beta_{\delta}=\frac{1}{2}\ln(t_{+}/t_{-})=\operatorname{arctanh}(J_{\delta}/\Delta_{\delta})$, which in the weak-exchange limit $J_{\delta}\ll\Delta_{\delta}$ reduces to $\beta_{\delta}\simeq J_{\delta}/\Delta_{\delta}$.

\paragraph{The unitary factor: mismatched spin frames.}

We now consider the microscopic origin of the unitary factor
$U_{\delta}$. The intermediate orbital $|d_{\delta}\rangle$ couples to
the two neighboring sites $i$ and $j=i+\delta$. We describe this coupling
by the hybridization Hamiltonian
\begin{equation}
H_{\mathrm{hyb}}
=
v_{\delta}
\left(
c_i^\dagger \bm d_{\delta}
+
c_j^\dagger \bm d_{\delta}
+
\mathrm{H.c.}
\right).
\label{eq:Hhyb}
\end{equation}

Here $c_i=(c_{i\uparrow},c_{i\downarrow})^T$ denotes the electron
annihilation spinor on magnetic site $i$, with
$c_i^\dagger=(c_{i\uparrow}^\dagger,c_{i\downarrow}^\dagger)$ its
creation spinor, and analogously for site $j$. Since the intermediate orbital lies outside the low-energy sector, it is
only virtually occupied. Eliminating this orbital to second order in
$v_\delta$ generates an effective hopping between sites $i$ and $j$.
The electron first hops from $i$ to the intermediate orbital with
amplitude $v_\delta$, propagates virtually through this orbital, and
then hops to $j$ with another factor $v_\delta$. At low energies, the
virtual propagation through the intermediate orbital is described by
$h_{d,\delta}^{-1}$. The resulting effective hopping matrix is therefore
proportional to
\begin{equation}
T_\delta
=
v_\delta^2 h_{d,\delta}^{-1}.
\label{eq:Teff_simple}
\end{equation}

This follows from the low-energy equation for the intermediate orbital,
$h_{d,\delta}\bm d_\delta \simeq
-v_\delta(c_i+c_j)$, which gives
$\bm d_\delta\simeq
-v_\delta h_{d,\delta}^{-1}(c_i+c_j)$.

The local magnetic moments at sites $i$ and $j$, as well as the magnetic
environment of the intermediate orbital, can define different spin
quantization axes. We denote by $R_i$, $R_j$, and $R_{d,\delta}$ the
SU(2) rotations that relate these local spin frames to a common reference
frame. Accordingly,
\begin{equation}
c_i=R_i\widetilde c_i,
\qquad
c_j=R_j\widetilde c_j,
\qquad
\bm d_{\delta}
=
R_{d,\delta}\widetilde{\bm d}_{\delta}.
\end{equation}

Here the tilded spinors $\widetilde c_i$, $\widetilde c_j$, and
$\widetilde{\bm d}_{\delta}$ denote the corresponding fermionic
operators expressed in their respective local spin frames. Substituting these expressions into Eq.~\eqref{eq:Hhyb}, the hopping
between site $i$ and the intermediate orbital becomes
\begin{equation}
v_{\delta}c_i^\dagger\bm d_{\delta}
=
v_{\delta}\widetilde c_i^\dagger
R_i^\dagger R_{d,\delta}
\widetilde{\bm d}_{\delta}.
\end{equation}

Similarly, the hopping between the intermediate orbital and site $j$ is
\begin{equation}
v_\delta {\bm d}_\delta^\dagger c_j
=
v_\delta
\widetilde {\bm d}_\delta^\dagger
R_{d,\delta}^\dagger R_j
\widetilde c_j .
\end{equation}

The effective hopping from site $i$ to site $j$ therefore contains three
successive factors: the hopping from $i$ to the intermediate orbital,
the virtual propagation through this orbital, and the hopping from the
intermediate orbital to $j$. Thus,
\begin{equation}
T_\delta
=
v_\delta^2
\left(R_i^\dagger R_{d,\delta}\right)
h_{d,\delta}^{-1}
\left(R_{d,\delta}^\dagger R_j\right),
\label{eq:Teff_frames}
\end{equation}
where $h_{d,\delta}$ is written in the local spin frame of the
intermediate orbital.

Using $(\hat{\bm p}_{\delta}\cdot\bm{\sigma})^2=\sigma_0$, we have
\begin{equation}
\underbrace{
\left(
\Delta_\delta\sigma_0
-
J_\delta\hat{\bm p}_\delta\cdot\bm{\sigma}
\right)
}_{h_{d,\delta}}
\left(
\Delta_\delta\sigma_0
+
J_\delta\hat{\bm p}_\delta\cdot\bm{\sigma}
\right)
=
\left(
\Delta_\delta^2-J_\delta^2
\right)\sigma_0 .
\end{equation}
Dividing by $\Delta_\delta^2-J_\delta^2$, we obtain
\begin{equation}
h_{d,\delta}
\left[
\frac{
\Delta_\delta\sigma_0
+
J_\delta\hat{\bm p}_\delta\cdot\bm{\sigma}
}{
\Delta_\delta^2-J_\delta^2
}
\right]
=
\sigma_0 .
\end{equation}
Therefore, the term inside brackets is the inverse of
$h_{d,\delta}$,
\begin{equation}
h_{d,\delta}^{-1}
=
\frac{
\Delta_\delta\sigma_0
+
J_\delta\hat{\bm p}_\delta\cdot\bm{\sigma}
}{
\Delta_\delta^2-J_\delta^2
}.
\label{eq:hdinverse}
\end{equation}

To rewrite Eq.~(\ref{eq:hdinverse}) in the form used in the spin-bond
decomposition, we used  $\beta_\delta
=
\operatorname{arctanh}
\left(
\frac{J_\delta}{\Delta_\delta}
\right)$ and 
\begin{equation}
e^{\beta_\delta\hat{\bm p}_\delta\cdot\bm{\sigma}}
=
\cosh\beta_\delta\,\sigma_0
+
\sinh\beta_\delta\,
\hat{\bm p}_\delta\cdot\bm{\sigma}.
\end{equation}
Then, Eq.~(\ref{eq:hdinverse}) can be written as
\begin{equation}
h_{d,\delta}^{-1}
=
\frac{1}{\sqrt{\Delta_\delta^2-J_\delta^2}}
\exp\left(
\beta_\delta
\hat{\bm p}_\delta\cdot\bm{\sigma}
\right).
\label{eq:hdinverse_exp}
\end{equation}

Therefore,
\begin{equation}
T_\delta
=
t_\delta
\left(R_i^\dagger R_{d,\delta}\right)
\exp\left(
\beta_\delta
\hat{\bm p}_\delta\cdot\bm{\sigma}
\right)
\left(R_{d,\delta}^\dagger R_j\right),
\label{eq:Teff_exp}
\end{equation}
where
\begin{equation}
t_\delta
=
\frac{v_\delta^2}
{\sqrt{\Delta_\delta^2-J_\delta^2}}.
\end{equation}

To separate the relative spin rotation between sites $i$ and $j$, we
insert the identity $R_jR_j^\dagger=\sigma_0$ into Eq.~(\ref{eq:Teff_exp}). Thus,

\begin{equation}
T_\delta
=
t_\delta
\left(R_i^\dagger R_j\right)
\left[
R_j^\dagger R_{d,\delta}
\exp\left(
\beta_\delta\hat{\bm p}_\delta\cdot\bm{\sigma}
\right)
R_{d,\delta}^\dagger R_j
\right].
\label{eq:Teff_factored}
\end{equation}

The two factors in Eq.~\eqref{eq:Teff_factored} have precisely the
properties required by the polar decomposition. Since $R_i$ and $R_j$
are SU(2) rotations, the first factor is unitary. We therefore identify
\begin{equation}
U_\delta
=
R_i^\dagger R_j .
\label{eq:Umicro}
\end{equation}
As any SU(2) matrix, it can be written as
\begin{equation}
U_\delta
=
\exp\left(
i\alpha_\delta
\hat{\bm u}_\delta\cdot\bm{\sigma}
\right),
\label{eq:Umicro_exp}
\end{equation}
where $\alpha_\delta$ characterizes the relative rotation between the
local spin frames at sites $i$ and $j$, and $\hat{\bm u}_\delta$ is the
corresponding rotation axis. For a physical rotation by an angle $\theta_\delta$ about the axis
$\hat{\bm u}_\delta$, the SU(2) transformation is
\begin{equation}
U_\delta
=
\exp\left(
-\frac{i\theta_\delta}{2}
\hat{\bm u}_\delta\cdot\bm{\sigma}
\right).
\end{equation}
Comparison with
$U_\delta=\exp(i\alpha_\delta\hat{\bm u}_\delta\cdot\bm{\sigma})$
gives
\begin{equation}
\alpha_\delta=-\frac{\theta_\delta}{2}.
\end{equation}

The second factor in Eq.~\eqref{eq:Teff_factored} is Hermitian and
positive, since it is a unitary rotation of the positive matrix
$\exp(\beta_\delta\hat{\bm p}_\delta\cdot\bm{\sigma})$. We therefore
identify
\begin{equation}
P_\delta
=
R_j^\dagger R_{d,\delta}
\exp\left(
\beta_\delta
\hat{\bm p}_\delta\cdot\bm{\sigma}
\right)
R_{d,\delta}^\dagger R_j .
\label{eq:Pmicro}
\end{equation}

The unitary transformations in Eq.~\eqref{eq:Pmicro} only rotate the
spin axis. Thus, defining the rotated axis $\hat{\bm p}_\delta^{\,\prime}$
through
\begin{equation}
R_j^\dagger R_{d,\delta}
\left(
\hat{\bm p}_\delta\cdot\bm{\sigma}
\right)
R_{d,\delta}^\dagger R_j
=
\hat{\bm p}_\delta^{\,\prime}\cdot\bm{\sigma},
\end{equation}
the Hermitian factor takes the same form introduced in the spin-bond
theory,
\begin{equation}
P_\delta
=
\exp\left(
\beta_\delta
\hat{\bm p}_\delta^{\,\prime}\cdot\bm{\sigma}
\right).
\label{eq:Pmicro_exp}
\end{equation}

The rotated axis $\hat{\bm p}_\delta^{\,\prime}$ is the amplitude axis
seen in the frame of site $j$, and it is the one entering the spin-bond
theory. From here on we drop the prime and write $\hat{\bm p}_\delta$
for this rotated axis.

The effective hopping generated by the virtual process therefore takes
the polar form
\begin{equation}
T_\delta
=
t_\delta U_\delta P_\delta .
\label{eq:Tpolar_micro}
\end{equation}

Importantly, the ordering $U_\delta P_\delta$ follows directly from the
microscopic hopping process and does not require the two factors to
commute. In general, the rotation axis $\hat{\bm u}_\delta$ and the
amplitude axis $\hat{\bm p}_\delta$ need not be parallel, so
that $[U_\delta,P_\delta]\neq 0$.

\subsection{The effective bond and its mixed term}

Combining the results above, the effective hopping generated by the
virtual process takes the form
\begin{equation}
T_\delta^{\mathrm{eff}}
=
t_\delta
\exp\left(
i\alpha_\delta\hat{\bm u}_\delta\cdot\bm{\sigma}
\right)
\exp\left(
\beta_\delta\hat{\bm p}_\delta\cdot\bm{\sigma}
\right)
=
t_\delta U_\delta P_\delta .
\label{eq:final}
\end{equation}
Thus, the polar spin bond introduced in the main text emerges directly
from the microscopic model. Importantly, the ordering $U_\delta P_\delta$
follows from the virtual hopping process and does not require the two
factors to commute.

For weak $\alpha_\delta$ and $\beta_\delta$,
\begin{equation}
\frac{T_\delta^{\mathrm{eff}}}{t_\delta}
\simeq
\sigma_0
+i\alpha_\delta\hat{\bm u}_\delta\cdot\bm{\sigma}
+\beta_\delta\hat{\bm p}_\delta\cdot\bm{\sigma}
+i\alpha_\delta\beta_\delta
(\hat{\bm u}_\delta\cdot\hat{\bm p}_\delta)\sigma_0
-\alpha_\delta\beta_\delta
(\hat{\bm u}_\delta\times\hat{\bm p}_\delta)\cdot\bm{\sigma}.
\label{eq:expansion}
\end{equation}
The last term is the mixed contribution generated when the spin-rotation
and spin-amplitude axes are non-collinear. Equivalently,
\begin{equation}
[U_\delta,P_\delta]
\simeq
-2\alpha_\delta\beta_\delta
(\hat{\bm u}_\delta\times\hat{\bm p}_\delta)\cdot\bm{\sigma}.
\end{equation}

In the present microscopic model,
$\beta_\delta\simeq J_\delta/\Delta_\delta$ for
$J_\delta\ll\Delta_\delta$, while $\alpha_\delta$ measures the relative
rotation of the local spin frames. The mixed contribution therefore
scales as $\alpha_\delta\beta_\delta$, linking the non-commuting sector
directly to microscopic magnetic parameters. In the weak-exchange limit, the strength of the mixed contribution is therefore directly controlled by the product of the relative spin-frame rotation and the exchange-induced hopping asymmetry, $|\alpha_\delta\beta_\delta|
\simeq
\frac{|\theta_\delta|}{2}
\frac{|J_\delta|}{\Delta_\delta}$.

\section{Parity of the phase and amplitude sectors}
\label{app:parity_proofs}

The parity analysis below is restricted to reciprocal bond patterns satisfying,
\begin{equation}
T_{-\delta}=T_\delta^\dagger .
\label{eq:bond_reciprocity}
\end{equation}

No spatial inversion symmetry is required. When $[U_\delta,P_\delta]=0$, the reciprocity condition can
be written separately for the unitary and Hermitian factors as

\begin{equation}
\text{phase:}\qquad
U_{-\delta}=U_\delta^\dagger,
\qquad
\text{amplitude:}\qquad
P_{-\delta}=P_\delta .
\label{eq:pure_reciprocity}
\end{equation}
Thus, in the pure unitary sector the bond is conjugated under bond
reversal, whereas in the pure Hermitian sector it is unchanged. This
difference directly produces the opposite momentum parities derived
below.
 
\subsection*{Phase links give odd spin fields}
 
For a purely unitary bond
\(T_\delta=t_\delta e^{i\alpha_\delta\hat{\mathbf u}_\delta\cdot\boldsymbol\sigma}\),
reciprocity \(T_{-\delta}=T_\delta^\dagger\) gives, on pairing opposite bonds,
\begin{equation}
S(\mathbf k)
=
\sum_{\delta>0}
\left[
T_\delta e^{i\mathbf k\cdot\delta}
+
T_\delta^\dagger e^{-i\mathbf k\cdot\delta}
\right].
\end{equation}
With
\(e^{i\alpha_\delta\hat{\mathbf u}_\delta\cdot\boldsymbol\sigma}
=\cos\alpha_\delta\,\sigma_0+i\sin\alpha_\delta\,
\hat{\mathbf u}_\delta\cdot\boldsymbol\sigma\), the scalar part recombines into
cosines and the spin part into sines,
\begin{equation}
h_0(\mathbf k)
=
\sum_{\delta>0}
2t_\delta\cos\alpha_\delta\,\cos(\mathbf k\cdot\delta),
\qquad
\mathbf h(\mathbf k)
=
-\sum_{\delta>0}
2t_\delta\sin\alpha_\delta\,\hat{\mathbf u}_\delta\,
\sin(\mathbf k\cdot\delta).
\end{equation}
The sine form factor is a direct consequence of the conjugation in
Eq.~\eqref{eq:pure_reciprocity}: because the \(+\delta\) and \(-\delta\) links
carry conjugate matrices, their imaginary (spin) parts enter with opposite signs
on \(e^{\pm i\mathbf k\cdot\delta}\). Hence
\begin{equation}
\boxed{\;
h_0(-\mathbf k)=h_0(\mathbf k),
\qquad
\mathbf h(-\mathbf k)=-\,\mathbf h(\mathbf k),\;}
\end{equation}
 an odd-parity spin field, the spin-bond origin of \(p\)-wave textures and emergent non-relativistic spin--orbit coupling.

\subsection*{Amplitude links give even spin fields}
 
For a purely Hermitian bond
\(T_\delta=t_\delta e^{\beta_\delta\hat{\mathbf p}_\delta\cdot\boldsymbol\sigma}\),
the operator is its own conjugate, so
\(T_{-\delta}=T_\delta^\dagger=T_\delta\): the opposite bond carries the
\emph{same} matrix. Pairing opposite bonds then gives
\begin{equation}
S(\mathbf k)
=
\sum_{\delta>0}
T_\delta
\left[
e^{i\mathbf k\cdot\delta}+e^{-i\mathbf k\cdot\delta}
\right]
=
\sum_{\delta>0}
2\,T_\delta\,\cos(\mathbf k\cdot\delta),
\end{equation}
where the equal matrices on \(\pm\delta\) combine the two exponentials into a
single cosine. Expanding
\(e^{\beta_\delta\hat{\mathbf p}_\delta\cdot\boldsymbol\sigma}
=\cosh\beta_\delta\,\sigma_0+\sinh\beta_\delta\,
\hat{\mathbf p}_\delta\cdot\boldsymbol\sigma\) and separating
\(S=h_0\sigma_0+\mathbf h\cdot\boldsymbol\sigma\),
\begin{equation}
h_0(\mathbf k)
=
\sum_{\delta>0}
2t_\delta\cosh\beta_\delta\,\cos(\mathbf k\cdot\delta),
\qquad
\mathbf h(\mathbf k)
=
\sum_{\delta>0}
2t_\delta\sinh\beta_\delta\,\hat{\mathbf p}_\delta\,
\cos(\mathbf k\cdot\delta).
\end{equation}
Both pieces are built from the same even form factor
\(\cos(\mathbf k\cdot\delta)\), so
\begin{equation}
\boxed{\;
\mathbf h(-\mathbf k)=+\,\mathbf h(\mathbf k),\;}
\end{equation}
 an even-parity spin field, with altermagnetism as one bond-structured realization. Therefore, the parity is set by the behaviour of a bond under inversion: the phase factor
conjugates and yields odd (\(\sin\)) form factors, while the amplitude factor is
invariant and yields even (\(\cos\)) ones. Parity is therefore locked to the
polar character of the bond, independent of lattice details,
\begin{equation}
U_\delta\;\Rightarrow\;\mathbf h(-\mathbf k)=-\mathbf h(\mathbf k),
\qquad\qquad
P_\delta\;\Rightarrow\;\mathbf h(-\mathbf k)=+\mathbf h(\mathbf k).
\end{equation}

Within the reciprocal family considered here,
$T_{-\delta}=T_\delta^\dagger$;
lattices without a center of inversion are allowed and simply enrich
$\mathbf h(\mathbf k)$ with additional bond contributions, each still
obeying the same per-sector parity. When both factors are present and
their spin axes are non-collinear,
\(\hat{\mathbf u}_\delta\times\hat{\mathbf p}_\delta\neq0\), the field
\(\mathbf h(\mathbf k)\) acquires an odd and an even part simultaneously, giving
the mixed-parity texture of the main text.

When both polar factors are present, the reciprocity condition must be
applied to the full bond operator. In particular, when
$[U_\delta,P_\delta]\neq0$,
\begin{equation}
T_{-\delta}
=
T_\delta^\dagger
=
P_\delta U_\delta^\dagger .
\label{eq:mixed_reciprocity}
\end{equation}
Rewriting the reversed bond in polar form,
$T_{-\delta}=U_{-\delta}P_{-\delta}$, gives
\begin{equation}
U_{-\delta}=U_\delta^\dagger,
\qquad
P_{-\delta}
=
U_\delta P_\delta U_\delta^\dagger .
\label{eq:reverse_polar}
\end{equation}
Therefore, $P_{-\delta}=P_\delta$ only when the unitary and Hermitian
factors commute. When $[U_\delta,P_\delta]\neq0$, the full bond contains
both even- and odd-in-momentum spin contributions, giving the
mixed-parity texture discussed in the main text.

\section{Exact block factorization}
\label{app:block}

We derive the two-block form of Eq. 18. In the local spin frame introduced in the main text , the exchange is uniform, so the Bloch Hamiltonian Eq. 12 of the main text reads
\begin{equation}
H(\mathbf k)
=
\Delta\,(\tau_0\otimes\sigma_z)
-
(\tau_+\otimes S(\mathbf k))
-
(\tau_-\otimes S^\dagger(\mathbf k)),
\label{eq:app_H}
\end{equation}
with \(\tau_\pm=\tfrac12(\tau_x\pm i\tau_y)\). Reciprocity of the bonds makes
\(S(\mathbf k)\) Hermitian, \(S=S^\dagger\), so the two hopping terms combine into
a single \(\tau_x\) term,
\begin{equation}
-(\tau_+\otimes S)-(\tau_-\otimes S^\dagger)
=
-(\tau_++\tau_-)\otimes S
=
-\,\tau_x\otimes S(\mathbf k),
\end{equation}
and the Hamiltonian takes the compact form
\begin{equation}
H(\mathbf k)
=
\Delta\,(\tau_0\otimes\sigma_z)
-
\tau_x\otimes S(\mathbf k).
\label{eq:app_H_compact}
\end{equation}

Both sublattice operators in Eq.~\eqref{eq:app_H_compact} --- the identity
\(\tau_0\) carried by the exchange and the \(\tau_x\) carried by the hopping ---
are diagonal in the eigenbasis of \(\tau_x\),
\begin{equation}
\tau_x\,|s\rangle=s\,|s\rangle,
\qquad
|s\rangle=\frac{1}{\sqrt2}\big(|A\rangle+s\,|B\rangle\big),
\qquad
s=\pm1.
\end{equation}
Indeed \(\tau_0\) is the identity, so it acts as the number \(1\) on either state,
\(\tau_0|s\rangle=|s\rangle\); and \(\tau_x\) returns its eigenvalue,
\(\tau_x|s\rangle=s|s\rangle\). Because both are diagonal in the same basis, no
operator connects \(|s=+1\rangle\) to \(|s=-1\rangle\), and the \(4\times4\)
Hamiltonian splits into two decoupled \(2\times2\) spin blocks. Within the block
\(s\) we may replace each sublattice operator by its eigenvalue,
\begin{equation}
\tau_0\longrightarrow 1,
\qquad
\tau_x\longrightarrow s,
\end{equation}
so that Eq.~\eqref{eq:app_H_compact} becomes a purely spin Hamiltonian,
\begin{equation}
H_s(\mathbf k)
=
\Delta\,\sigma_z
-
s\,S(\mathbf k),
\qquad
s=\pm1.
\end{equation}

Inserting \(S(\mathbf k)=h_0(\mathbf k)\sigma_0+\mathbf h(\mathbf k)\cdot
\boldsymbol\sigma\) gives
\begin{equation}
H_s(\mathbf k)
=
-s\,h_0(\mathbf k)\,\sigma_0
+
\big[\Delta\hat{\mathbf z}-s\,\mathbf h(\mathbf k)\big]\cdot\boldsymbol\sigma,
\qquad
s=\pm1,
\end{equation}
which is Eq. 18 of the main text. Each block is a two-level Hamiltonian of the standard
form
\begin{equation}
H_s(\mathbf k)
=
\epsilon_s(\mathbf k)\,\sigma_0
+
\mathbf b_s(\mathbf k)\cdot\boldsymbol\sigma,
\qquad
\epsilon_s=-s\,h_0(\mathbf k),
\quad
\mathbf b_s=\Delta\hat{\mathbf z}-s\,\mathbf h(\mathbf k),
\end{equation}
whose eigenvalues are \(\epsilon_s\pm|\mathbf b_s|\) and whose eigenspinors point
along \(\pm\hat{\mathbf b}_s\). Explicitly,
\begin{equation}
E_{s,\pm}(\mathbf k)
=
-s\,h_0(\mathbf k)
\pm
\big|\Delta\hat{\mathbf z}-s\,\mathbf h(\mathbf k)\big|,
\end{equation}
\begin{equation}
\langle\mathbf s\rangle_{s,\pm}(\mathbf k)
=
\pm\,
\frac{\Delta\hat{\mathbf z}-s\,\mathbf h(\mathbf k)}
{\big|\Delta\hat{\mathbf z}-s\,\mathbf h(\mathbf k)\big|},
\end{equation}
which are the bands of Eq. 19 of the main text and the spin texture of Eq. 20 of the main text. The
scalar piece \(-s\,h_0\) shifts the two levels of a block rigidly and does not
affect the spin direction, while the spin points along the effective field
\(\mathbf b_s=\Delta\hat{\mathbf z}-s\,\mathbf h(\mathbf k)\), reflecting the competition
between the uniform exchange and the bond field.

\section{Explicit bond patterns and their spin--orbit textures}
\label{app:recipes}

Here we derive the entries of Table 1 from the main text. All forms follow from the
spin field of the unitary sector,
\begin{equation}
\mathbf h(\kk)
=
-\sum_{\delta>0}
2t_\delta\sin\alpha_\delta\,
\uhat_\delta\,
\sin(\kk\cdot\delta),
\label{eq:h_general_recipes}
\end{equation}
For the square lattice we take $\bm\delta_x=a\hat{\bm x}$ and 
$\bm\delta_y=a\hat{\bm y}$, and set the lattice constant $a=1$, \(\bm \delta=\hat{\mathbf x},\hat{\mathbf y}\). To
leading order in \(\kk\) (\(\sin k_i\to k_i\)),
\begin{equation}
\mathbf h(\kk)
=
-2t\big[
\sin\alpha_x\,\uhat_x\,\sin k_x
+
\sin\alpha_y\,\uhat_y\,\sin k_y
\big]
\;\xrightarrow{\ \sin k_i\to k_i\ }\;
-2t\big[
\sin\alpha_x\,\uhat_x\,k_x
+
\sin\alpha_y\,\uhat_y\,k_y
\big],
\label{eq:h_square}
\end{equation}
and the exact forms are recovered by \(k_i\to\sin k_i\). Writing
\(\mathbf h\cdot\boldsymbol\sigma\) and absorbing signs into
\(\lambda\equiv2t\sin\alpha\) gives each row directly.

\medskip
\noindent\emph{Rashba~\cite{Rashba1960}.} With \(\uhat_x=\hat{\mathbf y}\),
\(\uhat_y=-\hat{\mathbf x}\) and \(\alpha_x=\alpha_y=\alpha\),
Eq.~\eqref{eq:h_square} gives
\(\mathbf h=-2t\sin\alpha\,(\hat{\mathbf y}\,k_x-\hat{\mathbf x}\,k_y)\), i.e.
\(\mathbf h\cdot\boldsymbol\sigma=\lambda(k_x\sigma_y-k_y\sigma_x)\), the
tangential winding texture.

\noindent\emph{Dresselhaus~\cite{Dresselhaus1955}.} With \(\uhat_x=\hat{\mathbf x}\),
\(\uhat_y=-\hat{\mathbf y}\),
\(\mathbf h\cdot\boldsymbol\sigma=\lambda(k_x\sigma_x-k_y\sigma_y)\).

\noindent\emph{Radial (Weyl).} With \(\uhat_x=\hat{\mathbf x}\),
\(\uhat_y=\hat{\mathbf y}\),
\(\mathbf h\cdot\boldsymbol\sigma=\lambda(k_x\sigma_x+k_y\sigma_y)\), a radial
(hedgehog-like) in-plane texture.

\noindent\emph{Out-of-plane.} With \(\uhat_x=\zhat\),
\(\mathbf h\cdot\boldsymbol\sigma=\lambda\,k_x\sigma_z\), an Ising-like texture
polarized along \(\zhat\).

\noindent\emph{\(p\)-wave.} With \(\uhat=\zhat\) on both bonds but opposite phase
angles \(\alpha_x=+\alpha\), \(\alpha_y=-\alpha\), the two contributions give a single
odd component \(h_z(\kk)\propto\sin k_x-\sin k_y\), odd in \(\kk\).

\medskip

\noindent\emph{$\Gamma$-split (uniform Hermitian).}
With $P_x=P_y=e^{\beta\sigma_z}$, and $U_\delta=\sigma_0$, one obtains
\[
S(\mathbf{k})
=
2t e^{\beta\sigma_z}(\cos k_x+\cos k_y),
\]
and therefore $h_z(\mathbf{k})
=
2t\sinh\beta(\cos k_x+\cos k_y)$. This defines an even-parity compensated spin splitting that remains
finite at the Brillouin-zone centre.

\medskip

\noindent\emph{\(d\)-altermagnet (exact)~\cite{Smejkal2022b}.} A bond-structured Hermitian realization. With
\(P_x=e^{\beta\sigma_z}\), \(P_y=e^{-\beta\sigma_z}\) and \(U_\delta=\sigma_0\), pairing
opposite bonds gives, with no expansion in \(\kk\),
\begin{equation}
S(\kk)
=
2t\cosh\beta\,(\cos k_x+\cos k_y)\,\sigma_0
+
2t\sinh\beta\,(\cos k_x-\cos k_y)\,\sigma_z ,
\end{equation}
so \(h_z=2t\sinh\beta\,(\cos k_x-\cos k_y)\) is manifestly even in \(\kk\) --- the
\(d\)-wave altermagnetic form.

\section{Non-collinear exchange under canting}
\label{app:ncol}

The exact solution of the main text assumes a uniform collinear exchange.
A canting moment \(\MM\) added to the N\'eel field,
\(\mathbf h_{A,B}=\pm\Delta\Nhat+\MM\), modifies this. Aligning each sublattice with
its local rotation (\(R_A\Nhat=R_B(-\Nhat)=\zhat\)), the exchange becomes
\begin{equation}
H'_{\rm ex}
=
\tau_0\otimes\big(\Delta\sigma_z+\MM_+\!\cdot\!\bsig\big)
+
\tau_z\otimes\big(\MM_-\!\cdot\!\bsig\big),
\qquad
\MM_\pm=\tfrac12(R_A\MM\pm R_B\MM).
\label{eq:canted_exchange}
\end{equation}
The uniform part \(\tau_0\otimes(\Delta\sigma_z+\MM_+\!\cdot\!\bsig)\) preserves the
two-block structure. The staggered part \(\tau_z\otimes\MM_-\!\cdot\!\bsig\) does
not: \(\tau_z\) does not commute with the \(\tau_x\) hopping, so it couples the
\(s=\pm1\) blocks and removes the exact parity factorization, just as a staggered
exchange does in App.~\ref{app:block}. We do not develop the canted case here. For small \(\MM_-\) it can be reached perturbatively by a Schrieffer--Wolff projection~\cite{SchriefferWolff1966} onto the lower block, which
yields the canting-induced corrections to the effective spin--orbit field; for
larger canting a numerical treatment is required. We note one caveat of the
projection: a single-sublattice Schrieffer--Wolff transformation captures the
ground-state spin direction but not the \(O(S)\) inter-sublattice splitting, so the
exact treatment of the main text is preferred whenever \(S\) is Hermitian.
In all cases the parity classification of the main text is the \(\MM_-\to0\) limit,
recovered continuously as the canting is switched off.

\section{The spin-bond generator}
\label{app:generator}

A complementary, more geometric view of the polar decomposition follows from the
spin-bond generator
\begin{equation}
T_\delta=e^{\mathcal G_\delta},\qquad \mathcal G_\delta\equiv\log T_\delta .
\label{eq:gen_def}
\end{equation}
We use the principal branch of the logarithm and assume weak spin-dependent
couplings. The spin-independent factor \(t_\delta e^{i\phi_\delta}\) is omitted
(it would add \((\log t_\delta+i\phi_\delta)\sigma_0\) to the generator), so the
spin-dependent part is traceless and Pauli-valued.

For a mixed link \(T_\delta=U_\delta P_\delta=
e^{i\alpha_\delta\uhat_\delta\cdot\bsig}\,e^{\beta_\delta\phat_\delta\cdot\bsig}\),
the Baker--Campbell--Hausdorff formula
\begin{equation}
\log(e^Ae^B)=A+B+\tfrac12[A,B]+\tfrac1{12}\big([A,[A,B]]+[B,[B,A]]\big)+\cdots
\end{equation}
with \(A=i\alpha_\delta\uhat_\delta\cdot\bsig\),
\(B=\beta_\delta\phat_\delta\cdot\bsig\), and
\([\mathbf a\cdot\bsig,\mathbf b\cdot\bsig]=2i(\mathbf a\times\mathbf b)\cdot\bsig\),
gives
\begin{align}
\mathcal G_\delta
&=
i\alpha_\delta\uhat_\delta\cdot\bsig
+\beta_\delta\phat_\delta\cdot\bsig
-\alpha_\delta\beta_\delta(\uhat_\delta\times\phat_\delta)\cdot\bsig
\nonumber\\
&\quad
-\frac{\alpha_\delta^2\beta_\delta}{3}\,\phat^{(u)}_{\delta,\perp}\cdot\bsig
+\frac{i\alpha_\delta\beta_\delta^2}{3}\,\uhat^{(p)}_{\delta,\perp}\cdot\bsig
+\cdots ,
\label{eq:G_expansion}
\end{align}
where the transverse projections are
\begin{equation}
\phat^{(u)}_{\delta,\perp}=\phat_\delta-(\uhat_\delta\cdot\phat_\delta)\uhat_\delta,
\qquad
\uhat^{(p)}_{\delta,\perp}=\uhat_\delta-(\uhat_\delta\cdot\phat_\delta)\phat_\delta .
\end{equation}

The first term is the anti-Hermitian generator of a spin-dependent phase; the
second is the Hermitian generator of a spin-dependent amplitude. The conversion
between them is bidirectional. The leading channel is phase-to-amplitude: the
mixed term
\begin{equation}
\mathcal G_\delta^{\rm amp}
=-\alpha_\delta\beta_\delta\,(\uhat_\delta\times\phat_\delta)\cdot\bsig
\label{eq:gen_amp_app}
\end{equation}
is \emph{Hermitian}, so a spin phase and a spin amplitude combine to produce a new
spin \emph{amplitude} along the emergent axis \(\uhat_\delta\times\phat_\delta\); it
vanishes when \(\uhat_\delta\parallel\phat_\delta\). The reciprocal channel,
amplitude-to-phase, appears one order higher as the \emph{anti-Hermitian} term
\begin{equation}
\mathcal G_\delta^{\rm ph}
=\frac{i\,\alpha_\delta\beta_\delta^2}{3}\,
\uhat^{(p)}_{\delta,\perp}\cdot\bsig ,
\label{eq:gen_ph_app}
\end{equation}
a new spin \emph{phase} along \(\uhat^{(p)}_{\delta,\perp}\), the component of
\(\uhat_\delta\) transverse to \(\phat_\delta\). The remaining higher-order term,
\(-\tfrac{\alpha_\delta^2\beta_\delta}{3}\phat^{(u)}_{\delta,\perp}\cdot\bsig\), is
Hermitian and corrects the amplitude along \(\phat^{(u)}_{\delta,\perp}\). Thus
non-commutativity converts phase into amplitude at order
\(\alpha_\delta\beta_\delta\) and, reciprocally, amplitude into phase at order
\(\alpha_\delta\beta_\delta^2\), the phase-to-amplitude channel being the leading
one. Each correction projects one bond axis onto the subspace transverse to the
other: a transverse projection of non-commutative origin, not a dipole
interaction.

Decomposing the generator into Hermitian and anti-Hermitian vector parts,
\(\mathcal G_\delta=(\mathbf b_\delta+i\mathbf a_\delta)\cdot\bsig\), the expansion
Eq.~\eqref{eq:G_expansion} gives
\begin{align}
\mathbf a_\delta
&=\alpha_\delta\uhat_\delta
+\frac{\alpha_\delta\beta_\delta^2}{3}\uhat^{(p)}_{\delta,\perp}+\cdots ,\\
\mathbf b_\delta
&=\beta_\delta\phat_\delta
-\alpha_\delta\beta_\delta(\uhat_\delta\times\phat_\delta)
-\frac{\alpha_\delta^2\beta_\delta}{3}\phat^{(u)}_{\delta,\perp}+\cdots .
\end{align}
The phase field \(\mathbf a_\delta\) and amplitude field \(\mathbf b_\delta\) are
independent only in the commuting limit; when the axes do not commute, each induces
corrections to the other. From \(\tfrac12\mathrm{Tr}\,\mathcal G_\delta^2=
(\mathbf b_\delta+i\mathbf a_\delta)^2=I_{1,\delta}+iI_{2,\delta}\) one obtains the
local invariants
\begin{equation}
I_{1,\delta}=\mathbf b_\delta^2-\mathbf a_\delta^2,\qquad
I_{2,\delta}=2\,\mathbf a_\delta\cdot\mathbf b_\delta ,
\end{equation}
where \(I_{1,\delta}\) measures the balance between amplitude and phase content
(positive: Hermitian-dominated; negative: unitary-dominated) and \(I_{2,\delta}\)
their longitudinal coupling. They are diagnostics of the local character of a
bond, not response coefficients.

The Hermitian and anti-Hermitian components of the same bond also
provide a local measure of phase--amplitude mixing. We define
\begin{equation}
\chi_{\mathrm{mix},\delta}
=
\left|
\bm a_\delta\times\bm b_\delta
\right|.
\label{eq:chi_mix_generator}
\end{equation}
To lowest order,
\begin{equation}
\bm a_\delta\times\bm b_\delta
=
\alpha_\delta\beta_\delta
\left(
\hat{\bm u}_\delta\times\hat{\bm p}_\delta
\right)
+\cdots ,
\end{equation}
so that
\begin{equation}
\chi_{\mathrm{mix},\delta}
\simeq
\alpha_\delta\beta_\delta
\left|
\hat{\bm u}_\delta\times\hat{\bm p}_\delta
\right|.
\end{equation}
This quantity vanishes when the local spin-phase and spin-amplitude
axes are parallel and is maximal when they are orthogonal. The same
factor $|\hat{\bm u}_\delta\times\hat{\bm p}_\delta|$ controls both
the leading local mixed term and the even-in-momentum transverse
polarization of the main text.

Independently of this local phase--amplitude mixing, different bonds
may themselves fail to commute. Writing
$\bm g_\delta=\bm b_\delta+i\bm a_\delta$, one finds, for
$\delta\neq\delta'$,
\begin{equation}
[G_\delta,G_{\delta'}]
=
2i
\left(
\bm g_\delta\times\bm g_{\delta'}
\right)\cdot\bm\sigma .
\label{eq:interbond_commutator}
\end{equation}
This relation characterizes non-commutativity between different
spin-dependent bonds and should be distinguished from the local
condition $[U_\delta,P_\delta]\neq0$.

The generator also defines a discrete spin-bond curvature. Around a
plaquette,
\begin{equation}
W_\square
=
e^{G_x}e^{G_y}e^{-G_x}e^{-G_y}
\simeq
e^{[G_x,G_y]},
\qquad
F^{\rm bond}_{xy}
\equiv
[G_x,G_y]
=
2i(\bm g_x\times\bm g_y)\cdot\bm\sigma .
\end{equation}
This quantity characterizes the inter-bond non-commutativity of the
spin-bond connection: it vanishes for commuting generators and is
finite when the bond generators are non-parallel.

\section{Quantitative Estimations for Synthetic SOC}

In this section we make use of the spin-bond formalism for quantifying the emergent synthetic SOC and its mechanical tunability. To understand how synthetic spin-orbit coupling emerges without relativistic effects, we begin by recalling the exact polar decomposition of the spin-dependent bond operator ($T_\delta$) into its unitary phase ($U_\delta$) and Hermitian amplitude ($P_\delta$) sectors:
\begin{equation}
T_\delta = U_\delta P_\delta \label{eq:T_polar}
\end{equation}
The dimensionless spin-dependent coupling constants $\alpha$ and $\beta$ are intrinsically small ($\alpha, \beta \ll 1$), as the secondary energy scales they parameterize ($t\alpha$ and $t\beta$) are comparable in
magnitude to relativistic spin-orbit interactions, and are therefore much smaller than the primary
electronic bandwidth \cite{Smejkal2022b, Krempasky2024}. This separation of energy scales allows us to perform a first-order Taylor expansion on the exponential operators, separating the spin-independent identity from the spin-dependent perturbations:
\begin{align}
U_\delta &= \exp(i \alpha \hat{\mathbf{u}} \cdot \boldsymbol{\sigma}) \approx \mathbb{I} + i \alpha (\hat{\mathbf{u}} \cdot \boldsymbol{\sigma}) \label{eq:U_expand} \\
P_\delta &= \exp(\beta \hat{\mathbf{p}} \cdot \boldsymbol{\sigma}) \approx \mathbb{I} + \beta (\hat{\mathbf{p}} \cdot \boldsymbol{\sigma}) \label{eq:P_expand}
\end{align}
Substituting these expansions back into the polar decomposition, we multiply both sectors to construct the bond operator up to second order in the coupling constants:
\begin{equation}
T_\delta \approx \mathbb{I} + i \alpha (\hat{\mathbf{u}} \cdot \boldsymbol{\sigma}) + \beta (\hat{\mathbf{p}} \cdot \boldsymbol{\sigma}) + i \alpha \beta (\hat{\mathbf{u}} \cdot \boldsymbol{\sigma})(\hat{\mathbf{p}} \cdot \boldsymbol{\sigma}) \label{eq:T_expand}
\end{equation}
Applying now the identity, $(\hat{\mathbf{u}} \cdot \boldsymbol{\sigma})(\hat{\mathbf{p}} \cdot \boldsymbol{\sigma}) = (\hat{\mathbf{u}} \cdot \hat{\mathbf{p}})\mathbb{I} + i (\hat{\mathbf{u}} \times \hat{\mathbf{p}}) \cdot \boldsymbol{\sigma}$, the mixed term expands as:
\begin{equation}
i \alpha \beta \left[ (\hat{\mathbf{u}} \cdot \hat{\mathbf{p}})\mathbb{I} + i (\hat{\mathbf{u}} \times \hat{\mathbf{p}}) \cdot \boldsymbol{\sigma} \right] = i \alpha \beta (\hat{\mathbf{u}} \cdot \hat{\mathbf{p}})\mathbb{I} - \alpha \beta (\hat{\mathbf{u}} \times \hat{\mathbf{p}}) \cdot \boldsymbol{\sigma} \label{eq:mixed_expand}
\end{equation}

The transverse component in Eq.~\eqref{eq:mixed_expand} is $- \alpha \beta (\hat{\mathbf{u}} \times \hat{\mathbf{p}}) \cdot \boldsymbol{\sigma}$. As the full tight-binding Hamiltonian is constructed by scaling the dimensionless bond operator $T_\delta$ by the scalar hopping energy $t$ (i.e., $\mathcal{H}_{\text{hop}} \propto t T_\delta$), this transverse term acts as an effective, momentum-dependent magnetic field. Its amplitude is thus given by the product of the hopping energy and the cross-product magnitude, with a factor of two from pairing the $\pm\delta$ bonds:
\begin{equation}
\lambda_{\text{SOC}} = 2 t \alpha \beta \, |\hat{\mathbf{u}} \times \hat{\mathbf{p}}| \label{eq:lambda_SOC}
\end{equation}
Assuming orthogonal quantization axes ($\hat{\mathbf{u}} \perp \hat{\mathbf{p}}$) to maximize the amplitude, which also requires $[U_\delta,P_\delta]\neq0$, it reduces to a synthetic SOC energy of $\lambda_{\text{SOC}} =2 t \alpha \beta$.
\paragraph{Band Splitting:}the macroscopic altermagnetic spin splitting $\Delta_{\text{alt}}$, can be measured via ARPES. However, to derive this splitting theoretically, we assume the pure altermagnetic limit ($\alpha \to 0$) with $d$-wave symmetry ($\beta_x = +\beta$, $\beta_y = -\beta$) and collinear polarization ($\hat{\mathbf{p}} = \hat{\mathbf{z}}$):
\begin{equation}
T_x = \exp(+\beta \sigma_z) \quad \text{and} \quad T_y = \exp(-\beta \sigma_z) \label{eq:T_alt}
\end{equation}
The kinetic tight-binding Hamiltonian is:
\begin{equation}
\mathcal{H}(\mathbf{k}) = -t \left( T_x e^{i k_x} + T_x^\dagger e^{-i k_x} \right) - t \left( T_y e^{i k_y} + T_y^\dagger e^{-i k_y} \right) \label{eq:H_alt_k}
\end{equation}
As the links are purely Hermitian ($T = T^\dagger$), this simplifies to:
\begin{equation}
\mathcal{H}(\mathbf{k}) = -2t \left[ \exp(+\beta \sigma_z) \cos k_x + \exp(-\beta \sigma_z) \cos k_y \right] \label{eq:H_alt_cos}
\end{equation}
Expressing the exponential as $\exp(\pm \beta \sigma_z) = \cosh\beta \mathbb{I} \pm \sinh\beta \sigma_z$, and isolating the purely spin-dependent term:
\begin{equation}
\mathcal{H}_{\text{alt}}(\mathbf{k}) = - 2t \sinh\beta (\cos k_x - \cos k_y) \sigma_z \label{eq:H_alt_final}
\end{equation}
Evaluating the geometric factor at the Brillouin zone boundaries (e.g., $\mathbf{k} = (\pi, 0)$), the maximum energy splitting is $\Delta_{\text{alt}} = 8t \sinh\beta$. For moderate $\beta$, we apply Taylor expansion ($\sinh\beta \approx \beta$) to obtain:
\begin{equation}
\Delta_{\text{alt}} \approx 8 t \beta \label{eq:Delta_alt}
\end{equation}
Extracting $\beta \approx \Delta_{\text{alt}} / 8t$ and substituting analytically into Eq.~\eqref{eq:lambda_SOC}, the hopping integrals cancel, yielding the direct scaling relation:
\begin{equation}
\lambda_{\text{SOC}} \approx \frac{\alpha}{4} \Delta_{\text{alt}} \label{eq:lambda_scaling}
\end{equation}
\paragraph{Tunability of ($\delta \lambda_{\text{SOC}}$) via Strain:} 
Here we propose to tune the synthetic SOC via mechanical deformation. For instance by elastic strain, $\varepsilon = \delta r / r$ (where $r$ is the equilibrium interatomic distance) one can modulate the hopping integral $t$. Since orbital overlap typically decays as a power law with distance ($t \propto r^{-n}$)~\cite{Harrison1989}, the differential variation of the hopping energy is:
\begin{equation}
\frac{\delta t}{t} = -n \varepsilon \label{eq:harrison}
\end{equation}
Therefore, as the synthetic SOC scale is proportional to $t$, it deterministically inherits this mechanical response:
\begin{equation}
\delta \lambda_{\text{SOC}} = n \varepsilon \lambda_{\text{SOC}} =2 n \varepsilon t \alpha \beta \label{eq:delta_lambda}
\end{equation}

\section{Impact on Electric Dipole Spin Resonance (EDSR): Rabi Frequency}
Recent literature has increasingly highlighted altermagnets as highly promising platforms for magnetic-field-free spin qubits and scalable quantum architectures~\cite{AbadilloUriel2026,Kirczenow2026,Vakili2026,VosoughiNia2025,Steinacker2025}.  Traditionally, achieving fast spin control via Electric Dipole Spin Resonance (EDSR) requires engineering magnetic gradients using micromagnets. However, these micromagnets set some challenges when trying to scale up a spin-based quantum processor. Altermagnets offer a paradigm shift by theoretically protecting spin qubits at zero net magnetic field~\cite{AbadilloUriel2026,Kirczenow2026,Vakili2026,VosoughiNia2025}. Hence, to translate these conceptual proposals into functional hardware, a microscopic mechanism to dynamically drive spin manipulations without reintroducing these external magnetic components is required. Due to the fact that our synthetic SOC emerges purely from the rigid lattice geometry, it enables to be tuned on the fly via applied gate voltages using localized piezoelectric strain.
For an electron confined in a quantum dot of size $x_0$ with an orbital level spacing $\Delta E_{\text{orb}}$, an alternating electric field $E_{\text{ac}}$ drives spin rotations via EDSR~\cite{Golovach2006,Nowack2007}. The resulting Rabi frequency is determined by the ratio of the SOC gradient to the orbital spacing:
\begin{equation}
f_{\text{Rabi}} = \frac{e E_{\text{ac}} x_0}{h} \left( \frac{\delta \lambda_{\text{SOC}}}{\Delta E_{\text{orb}}} \right).\label{eq:f_rabi_gen}
\end{equation}
Substituting the strain modulation Eq.~\eqref{eq:delta_lambda}, we obtain:
\begin{equation}
f_{\text{Rabi}} = \frac{e E_{\text{ac}} x_0}{h \Delta E_{\text{orb}}} \left( n \varepsilon t \alpha \beta \right). \label{eq:f_rabi_final}
\end{equation}

Numerical Case Study: Manganese telluride (MnTe) is an intrinsic semiconductor and a confirmed altermagnet~\cite{Smejkal2022b,Krempasky2024}. Indeed, recent ARPES measurements report a band splitting of $\Delta_{\text{alt}} \approx 150$ meV~\cite{Krempasky2024}. In collinear MnTe, $\alpha$ is of relativistic origin (Te spin--orbit
coupling), which enters the same factor $U_\delta$. Assuming a modest unitary phase magnitude of $\alpha = 0.05$, the available synthetic SOC is:
\begin{equation}
\lambda_{\text{SOC}} = \frac{0.05}{4} (150 \text{ meV}) \approx 1.87 \text{ meV}. \label{eq:num_lambda}
\end{equation}
Applying a $1\%$ strain ($\varepsilon = 0.01$) and assuming an orbital decay exponent $n=2$:
\begin{equation}
\delta \lambda_{\text{SOC}} = 2 \times 0.01 \times 1.87 \text{ meV} \approx 37.5\ \mu\text{eV}. \label{eq:num_delta}
\end{equation}
Using standard semiconductor quantum dot parameters ($E_{\text{ac}} = 10^4$ V/m, $x_0 = 20$ nm, $\Delta E_{\text{orb}} = 1$ meV)~\cite{NadjPerge2010}, the AC driving energy is $e E_{\text{ac}} x_0 \approx 0.2$ meV. Hence, the resulting Rabi frequency is:
\begin{equation}
f_{\text{Rabi}} = \frac{1}{h} \left[ 0.2 \text{ meV} \times \left( \frac{0.0375 \text{ meV}}{1 \text{ meV}} \right) \right] \approx 1.81 \text{ GHz}. \label{eq:num_f_rabi}
\end{equation}
The drive frequency is set by the local qubit splitting at the dot and not by the $\Delta_{\text{alt}}$ value. This establishes that an altermagnetic spin qubit defined in MnTe can achieve sub-nanosecond gate times. Furthermore, because the tunable part of the synthetic SOC scales with the applied
strain, the drive switches off at $\varepsilon=0$, leaving only the static residual
$\lambda_{\text{SOC}}=2t\alpha\beta$. Whether this reduces sensitivity to charge noise
requires a separate analysis of the operating point.

\bibliographystyle{naturemag}

\begin{thebibliography}{10}
\expandafter\ifx\csname url\endcsname\relax
  \def\url#1{\texttt{#1}}\fi
\expandafter\ifx\csname urlprefix\endcsname\relax\def\urlprefix{URL }\fi
\providecommand{\bibinfo}[2]{#2}
\providecommand{\eprint}[2][]{\url{#2}}

\bibitem{Smejkal2022a}
\bibinfo{author}{{\v{S}}mejkal, L.}, \bibinfo{author}{Sinova, J.} \& \bibinfo{author}{Jungwirth, T.}
\newblock \bibinfo{title}{Beyond conventional ferromagnetism and antiferromagnetism: A phase with nonrelativistic spin and lattice rotation symmetry}.
\newblock \emph{\bibinfo{journal}{Phys.\ Rev.\ X}} \textbf{\bibinfo{volume}{12}}, \bibinfo{pages}{031042} (\bibinfo{year}{2022}).

\bibitem{Smejkal2022b}
\bibinfo{author}{{\v{S}}mejkal, L.}, \bibinfo{author}{Sinova, J.} \& \bibinfo{author}{Jungwirth, T.}
\newblock \bibinfo{title}{Emerging research landscape of altermagnetism}.
\newblock \emph{\bibinfo{journal}{Phys.\ Rev.\ X}} \textbf{\bibinfo{volume}{12}}, \bibinfo{pages}{040501} (\bibinfo{year}{2022}).

\bibitem{Mazin2022}
\bibinfo{author}{Mazin, I.~I.} \emph{et~al.}
\newblock \bibinfo{title}{Altermagnetism?a new punch line of fundamental magnetism}.
\newblock \emph{\bibinfo{journal}{Proc.\ Natl.\ Acad.\ Sci.}} \textbf{\bibinfo{volume}{118}}, \bibinfo{pages}{e2108924118} (\bibinfo{year}{2021}).

\bibitem{Smejkal2020}
\bibinfo{author}{{\v{S}}mejkal, L.} \emph{et~al.}
\newblock \bibinfo{title}{Crystal time-reversal symmetry breaking and spontaneous hall effect in collinear antiferromagnets}.
\newblock \emph{\bibinfo{journal}{Sci.\ Adv.}} \textbf{\bibinfo{volume}{6}}, \bibinfo{pages}{eaaz8809} (\bibinfo{year}{2020}).

\bibitem{Hayami2019}
\bibinfo{author}{Hayami, S.}, \bibinfo{author}{Yanagi, Y.} \& \bibinfo{author}{Kusunose, H.}
\newblock \bibinfo{title}{Momentum-dependent spin splitting by collinear antiferromagnetic ordering}.
\newblock \emph{\bibinfo{journal}{J.\ Phys.\ Soc.\ Jpn.}} \textbf{\bibinfo{volume}{88}}, \bibinfo{pages}{123702} (\bibinfo{year}{2019}).

\bibitem{Ahn2019}
\bibinfo{author}{Ahn, K.-H.}, \bibinfo{author}{Hariki, A.}, \bibinfo{author}{Lee, K.-W.} \& \bibinfo{author}{Kune\v{s}, J.}
\newblock \bibinfo{title}{Antiferromagnetism in {R}u{O}$_2$ as $d$-wave {P}omeranchuk instability}.
\newblock \emph{\bibinfo{journal}{Phys. Rev. B}} \textbf{\bibinfo{volume}{99}}, \bibinfo{pages}{184432} (\bibinfo{year}{2019}).

\bibitem{Hellenes2024}
\bibinfo{author}{Hellenes, A.~B.} \emph{et~al.}
\newblock \bibinfo{title}{P-wave magnets}.
\newblock \emph{\bibinfo{journal}{arXiv:2309.01607}}  (\bibinfo{year}{2024}).

\bibitem{Mitscherling2026}
\bibinfo{author}{Mitscherling, J.}, \bibinfo{author}{Priessnitz, J.}, \bibinfo{author}{Geschner, C.~K.} \& \bibinfo{author}{{\v{S}}mejkal, L.}
\newblock \bibinfo{title}{Microscopic origin of \(p\)-wave magnetism}.
\newblock \emph{\bibinfo{journal}{arXiv:2603.09736}}  (\bibinfo{year}{2026}).

\bibitem{Bhowal2024}
\bibinfo{author}{Bhowal, S.} \& \bibinfo{author}{Spaldin, N.~A.}
\newblock \bibinfo{title}{Ferroically ordered magnetic octupoles in \(d\)-wave altermagnets}.
\newblock \emph{\bibinfo{journal}{Phys.\ Rev.\ X}} \textbf{\bibinfo{volume}{14}}, \bibinfo{pages}{011019} (\bibinfo{year}{2024}).

\bibitem{JaeschkeUbiergo2025}
\bibinfo{author}{Jaeschke-Ubiergo, R.} \emph{et~al.}
\newblock \bibinfo{title}{Atomic altermagnetism}.
\newblock \emph{\bibinfo{journal}{arXiv:2503.10797}}  (\bibinfo{year}{2025}).

\bibitem{Karetta2026}
\bibinfo{author}{Karetta, B.}, \bibinfo{author}{Verbeek, X.~H.}, \bibinfo{author}{Jaeschke-Ubiergo, R.}, \bibinfo{author}{{\v{S}}mejkal, L.} \& \bibinfo{author}{Sinova, J.}
\newblock \bibinfo{title}{Strain controlled \(g\)- to \(d\)-wave transition in altermagnetic {CrSb}}.
\newblock \emph{\bibinfo{journal}{Phys. Rev. B}} \textbf{\bibinfo{volume}{112}}, \bibinfo{pages}{094454} (\bibinfo{year}{2025}).

\bibitem{Yu2025}
\bibinfo{author}{Yu, Y.} \emph{et~al.}
\newblock \bibinfo{title}{Odd-parity magnetism driven by antiferromagnetic exchange}.
\newblock \emph{\bibinfo{journal}{Phys.\ Rev.\ Lett.}} \textbf{\bibinfo{volume}{135}}, \bibinfo{pages}{046701} (\bibinfo{year}{2025}).

\bibitem{Amin2024}
\bibinfo{author}{Amin, O.~J.} \emph{et~al.}
\newblock \bibinfo{title}{Nanoscale imaging and control of altermagnetism in mnte}.
\newblock \emph{\bibinfo{journal}{Nature}} \textbf{\bibinfo{volume}{636}}, \bibinfo{pages}{348} (\bibinfo{year}{2024}).

\bibitem{Zhou2025}
\bibinfo{author}{Zhou, Z.} \emph{et~al.}
\newblock \bibinfo{title}{Manipulation of the altermagnetic order in crsb via crystal symmetry}.
\newblock \emph{\bibinfo{journal}{Nature}} \textbf{\bibinfo{volume}{638}}, \bibinfo{pages}{645} (\bibinfo{year}{2025}).

\bibitem{Dil2009}
\bibinfo{author}{Dil, J.~H.}
\newblock \bibinfo{title}{Spin- and angle-resolved photoemission on non-magnetic low-dimensional systems}.
\newblock \emph{\bibinfo{journal}{J.\ Phys.\ Condens.\ Matter}} \textbf{\bibinfo{volume}{21}}, \bibinfo{pages}{403001} (\bibinfo{year}{2009}).

\bibitem{Dil2019}
\bibinfo{author}{Dil, J.~H.}
\newblock \bibinfo{title}{Spin- and angle-resolved photoemission on topological materials}.
\newblock \emph{\bibinfo{journal}{Electronic Structure}} \textbf{\bibinfo{volume}{1}}, \bibinfo{pages}{023001} (\bibinfo{year}{2019}).

\bibitem{Krempasky2024}
\bibinfo{author}{Kremask\'y, J.} \emph{et~al.}
\newblock \bibinfo{title}{Altermagnetic lifting of kramers spin degeneracy}.
\newblock \emph{\bibinfo{journal}{Nature}} \textbf{\bibinfo{volume}{626}}, \bibinfo{pages}{517--522} (\bibinfo{year}{2024}).

\bibitem{Reimers2024}
\bibinfo{author}{Reimers, S.} \emph{et~al.}
\newblock \bibinfo{title}{Direct observation of altermagnetic band splitting in crsb}.
\newblock \emph{\bibinfo{journal}{Nat.\ Commun.}} \textbf{\bibinfo{volume}{15}}, \bibinfo{pages}{2116} (\bibinfo{year}{2024}).

\bibitem{AbadilloUriel2026}
\bibinfo{author}{Abadillo-Uriel, J.~C.}, \bibinfo{author}{Maiani, A.}, \bibinfo{author}{Cortijo, A.}, \bibinfo{author}{Aguado, R.} \& \bibinfo{author}{Souto, R.~S.}
\newblock \bibinfo{title}{All-electrical dephasing-protected spin qubits in altermagnets}.
\newblock \emph{\bibinfo{journal}{arXiv:2606.26066}}  (\bibinfo{year}{2026}).

\bibitem{Kirczenow2026}
\bibinfo{author}{Kirczenow, G.}
\newblock \bibinfo{title}{Symmetry, disorder and transport through altermagnetic quantum dots and their antiferromagnetic twins}.
\newblock \emph{\bibinfo{journal}{Journal of Physics: Condensed Matter}} \textbf{\bibinfo{volume}{38}}, \bibinfo{pages}{175302} (\bibinfo{year}{2026}).

\bibitem{VosoughiNia2025}
\bibinfo{author}{Vosoughi-nia, S.} \& \bibinfo{author}{Nowak, M.~P.}
\newblock \bibinfo{title}{Altermon: a magnetic-field-free parity protected qubit based on a narrow altermagnet josephson junction}.
\newblock \emph{\bibinfo{journal}{arXiv:2510.18145}}  (\bibinfo{year}{2025}).

\bibitem{Steinacker2025}
\bibinfo{author}{Steinacker, P.} \emph{et~al.}
\newblock \bibinfo{title}{Industry-compatible silicon spin-qubit unit cells exceeding 99\% fidelity}.
\newblock \emph{\bibinfo{journal}{Nature}} \textbf{\bibinfo{volume}{646}}, \bibinfo{pages}{81} (\bibinfo{year}{2025}).

\bibitem{Vakili2026}
\bibinfo{author}{Vakili, H.}
\newblock \bibinfo{title}{Gate-controlled spin qubits in confined altermagnets}.
\newblock \emph{\bibinfo{journal}{arXiv preprint}}  (\bibinfo{year}{2026}).

\bibitem{Kulig2024}
\bibinfo{author}{Kulig, M.} \emph{et~al.}
\newblock \bibinfo{title}{The controlled rotation of entanglement in altermagnets}.
\newblock \emph{\bibinfo{journal}{arXiv preprint}}  (\bibinfo{year}{2024}).

\bibitem{Golovach2006}
\bibinfo{author}{Golovach, V.~N.}, \bibinfo{author}{Borhani, M.} \& \bibinfo{author}{Loss, D.}
\newblock \bibinfo{title}{Electric-dipole-induced spin resonance in quantum dots}.
\newblock \emph{\bibinfo{journal}{Phys.\ Rev.\ B}} \textbf{\bibinfo{volume}{74}}, \bibinfo{pages}{165319} (\bibinfo{year}{2006}).

\bibitem{Nowack2007}
\bibinfo{author}{Nowack, K.~C.}, \bibinfo{author}{Koppens, F. H.~L.}, \bibinfo{author}{Nazarov, Y.~V.} \& \bibinfo{author}{Vandersypen, L. M.~K.}
\newblock \bibinfo{title}{Coherent control of a single electron spin with electric fields}.
\newblock \emph{\bibinfo{journal}{Science}} \textbf{\bibinfo{volume}{318}}, \bibinfo{pages}{1430--1433} (\bibinfo{year}{2007}).

\bibitem{NadjPerge2010}
\bibinfo{author}{Nadj-Perge, S.}, \bibinfo{author}{Frolov, S.~M.}, \bibinfo{author}{Bakkers, E. P. A.~M.} \& \bibinfo{author}{Kouwenhoven, L.~P.}
\newblock \bibinfo{title}{Spin-orbit qubit in a semiconductor nanowire}.
\newblock \emph{\bibinfo{journal}{Nature}} \textbf{\bibinfo{volume}{468}}, \bibinfo{pages}{1084--1087} (\bibinfo{year}{2010}).

\bibitem{Sun2026}
\bibinfo{author}{Sun, W.} \& \bibinfo{author}{Jacob, Z.}
\newblock \bibinfo{title}{Correlated quantum dephasometry: Symmetry-resolved noise spectroscopy of two-dimensional superconductors and altermagnets}.
\newblock \emph{\bibinfo{journal}{arXiv preprint}}  (\bibinfo{year}{2026}).

\bibitem{Borhani2006}
\bibinfo{author}{Borhani, M.}, \bibinfo{author}{Golovach, V.~N.} \& \bibinfo{author}{Loss, D.}
\newblock \bibinfo{title}{Spin decay in a quantum dot coupled to a quantum point contact}.
\newblock \emph{\bibinfo{journal}{Phys. Rev. B}} \textbf{\bibinfo{volume}{73}}, \bibinfo{pages}{155311} (\bibinfo{year}{2006}).

\bibitem{Paladino2014}
\bibinfo{author}{Paladino, E.}, \bibinfo{author}{Galperin, Y.~M.}, \bibinfo{author}{Falci, G.} \& \bibinfo{author}{Altshuler, B.~L.}
\newblock \bibinfo{title}{1/f noise: Implications for solid-state quantum information}.
\newblock \emph{\bibinfo{journal}{Rev.\ Mod.\ Phys.}} \textbf{\bibinfo{volume}{86}}, \bibinfo{pages}{361} (\bibinfo{year}{2014}).

\bibitem{Yoneda2018}
\bibinfo{author}{Yoneda, J.} \emph{et~al.}
\newblock \bibinfo{title}{A quantum-dot spin qubit with coherence limited by charge noise and fidelity higher than 99.9\%}.
\newblock \emph{\bibinfo{journal}{Nat.\ Nanotechnol.}} \textbf{\bibinfo{volume}{13}}, \bibinfo{pages}{102--106} (\bibinfo{year}{2018}).

\bibitem{Burkard2023}
\bibinfo{author}{Burkard, G.}, \bibinfo{author}{Ladd, T.~D.}, \bibinfo{author}{Pan, A.}, \bibinfo{author}{Nichol, J.~M.} \& \bibinfo{author}{Petta, J.~R.}
\newblock \bibinfo{title}{Semiconductor spin qubits}.
\newblock \emph{\bibinfo{journal}{Rev. Mod. Phys.}} \textbf{\bibinfo{volume}{95}}, \bibinfo{pages}{025003} (\bibinfo{year}{2023}).

\bibitem{Ouassou2023}
\bibinfo{author}{Ouassou, J.~A.}, \bibinfo{author}{Brataas, A.} \& \bibinfo{author}{Linder, J.}
\newblock \bibinfo{title}{dc josephson effect in altermagnets}.
\newblock \emph{\bibinfo{journal}{Phys. Rev. Lett.}} \textbf{\bibinfo{volume}{131}}, \bibinfo{pages}{076003} (\bibinfo{year}{2023}).

\bibitem{Beenakker2023}
\bibinfo{author}{Beenakker, C. W.~J.} \& \bibinfo{author}{Vakhtel, T.}
\newblock \bibinfo{title}{Phase-shifted andreev levels in an altermagnet josephson junction}.
\newblock \emph{\bibinfo{journal}{Phys. Rev. B}} \textbf{\bibinfo{volume}{108}}, \bibinfo{pages}{075425} (\bibinfo{year}{2023}).

\bibitem{Giil2024}
\bibinfo{author}{Giil, H.~G.} \& \bibinfo{author}{Linder, J.}
\newblock \bibinfo{title}{Superconductor-altermagnet memory functionality without stray fields}.
\newblock \emph{\bibinfo{journal}{Phys. Rev. B}} \textbf{\bibinfo{volume}{109}}, \bibinfo{pages}{134511} (\bibinfo{year}{2024}).

\bibitem{Lu2024}
\bibinfo{author}{Lu, B.}, \bibinfo{author}{Maeda, K.}, \bibinfo{author}{Ito, H.}, \bibinfo{author}{Yada, K.} \& \bibinfo{author}{Tanaka, Y.}
\newblock \bibinfo{title}{$\varphi$ josephson junction induced by altermagnetism}.
\newblock \emph{\bibinfo{journal}{Phys. Rev. Lett.}} \textbf{\bibinfo{volume}{133}}, \bibinfo{pages}{226002} (\bibinfo{year}{2024}).

\bibitem{Banerjee2024}
\bibinfo{author}{Banerjee, S.} \& \bibinfo{author}{Scheurer, M.~S.}
\newblock \bibinfo{title}{Altermagnetic superconducting diode effect}.
\newblock \emph{\bibinfo{journal}{Phys. Rev. B}} \textbf{\bibinfo{volume}{110}}, \bibinfo{pages}{024503} (\bibinfo{year}{2024}).

\bibitem{Yuan2024}
\bibinfo{author}{Yuan, L.-D.}, \bibinfo{author}{Georgescu, A.~B.} \& \bibinfo{author}{Rondinelli, J.~M.}
\newblock \bibinfo{title}{Nonrelativistic spin splitting at the brillouin zone center in compensated magnets}.
\newblock \emph{\bibinfo{journal}{Phys.\ Rev.\ Lett.}} \textbf{\bibinfo{volume}{133}}, \bibinfo{pages}{216701} (\bibinfo{year}{2024}).

\bibitem{Han2024}
\bibinfo{author}{Han, L.} \emph{et~al.}
\newblock \bibinfo{title}{Observation of the antiferromagnetic spin hall effect}.
\newblock \emph{\bibinfo{journal}{Sci.\ Adv.}} \textbf{\bibinfo{volume}{10}}, \bibinfo{pages}{eadn0479} (\bibinfo{year}{2024}).

\bibitem{Chakraborty2025}
\bibinfo{author}{Chakraborty, A.} \emph{et~al.}
\newblock \bibinfo{title}{Highly efficient non-relativistic edelstein effect in nodal p-wave magnets}.
\newblock \emph{\bibinfo{journal}{Nat.\ Commun.}} \textbf{\bibinfo{volume}{16}}, \bibinfo{pages}{7270} (\bibinfo{year}{2025}).

\bibitem{Zhuang2026}
\bibinfo{author}{Zhuang, Z.-Y.} \emph{et~al.}
\newblock \bibinfo{title}{Odd-parity and non-abelian altermagnetism}.
\newblock \emph{\bibinfo{journal}{arXiv:2605.05205}}  (\bibinfo{year}{2026}).

\bibitem{Vandersypen2017}
\bibinfo{author}{Vandersypen, L. M.~K.} \emph{et~al.}
\newblock \bibinfo{title}{Interfacing spin qubits in quantum dots and donors?hot, dense, and coherent}.
\newblock \emph{\bibinfo{journal}{npj Quantum Inf.}} \textbf{\bibinfo{volume}{3}}, \bibinfo{pages}{34} (\bibinfo{year}{2017}).

\bibitem{Hetenyi2020}
\bibinfo{author}{Het\'enyi, B.}, \bibinfo{author}{Kloeffel, C.} \& \bibinfo{author}{Loss, D.}
\newblock \bibinfo{title}{Exchange interaction of hole-spin qubits in double quantum dots in highly anisotropic semiconductors}.
\newblock \emph{\bibinfo{journal}{Phys. Rev. Research}} \textbf{\bibinfo{volume}{2}}, \bibinfo{pages}{033036} (\bibinfo{year}{2020}).

\bibitem{PioroLadriere2008}
\bibinfo{author}{Pioro-Ladri{\`e}re, M.} \emph{et~al.}
\newblock \bibinfo{title}{Electrically driven single-electron spin resonance in a slanting zeeman field}.
\newblock \emph{\bibinfo{journal}{Nat. Phys.}} \textbf{\bibinfo{volume}{4}}, \bibinfo{pages}{776} (\bibinfo{year}{2008}).

\bibitem{Rashba1960}
\bibinfo{author}{Rashba, E.~I.}
\newblock \bibinfo{title}{Properties of semiconductors with an extremum loop. 1. cyclotron and combinational resonance in a magnetic field perpendicular to the plane of the loop}.
\newblock \emph{\bibinfo{journal}{Sov.\ Phys.\ Solid State}} \textbf{\bibinfo{volume}{2}}, \bibinfo{pages}{1109} (\bibinfo{year}{1960}).

\bibitem{Dresselhaus1955}
\bibinfo{author}{Dresselhaus, G.}
\newblock \bibinfo{title}{Spin-orbit coupling effects in zinc blende structures}.
\newblock \emph{\bibinfo{journal}{Phys.\ Rev.}} \textbf{\bibinfo{volume}{100}}, \bibinfo{pages}{580} (\bibinfo{year}{1955}).

\bibitem{SchriefferWolff1966}
\bibinfo{author}{Schrieffer, J.~R.} \& \bibinfo{author}{Wolff, P.~A.}
\newblock \bibinfo{title}{Relation between the anderson and kondo hamiltonians}.
\newblock \emph{\bibinfo{journal}{Phys.\ Rev.}} \textbf{\bibinfo{volume}{149}}, \bibinfo{pages}{491} (\bibinfo{year}{1966}).

\bibitem{Harrison1989}
\bibinfo{author}{Harrison, W.~A.}
\newblock \emph{\bibinfo{title}{Electronic Structure and the Properties of Solids: The Physics of the Chemical Bond}} (\bibinfo{publisher}{Dover Publications}, \bibinfo{year}{1989}).

\end{thebibliography}

\end{document}